\documentclass[twocolumn,showpacs,prc,aps,floatfix,hidelinks]{revtex4-1}

\usepackage{bm}
\usepackage{braket}
\usepackage{amssymb}
\usepackage{graphicx}
\usepackage{dcolumn}
\usepackage{amsmath}
\usepackage{amsfonts}
\usepackage{xcolor}
\usepackage{hyperref}

\usepackage{mathtools}
\mathtoolsset{showonlyrefs}

\begin{document}

\title{A dispersive optical model analysis of $^{208}$Pb generating a neutron-skin prediction beyond the mean field}
\author{M. C. Atkinson$^{1,2}$, M. H. Mahzoon$^2$, M. A. Keim$^2$, B. A. Bordelon$^2$, C. D. Pruitt$^3$, R. J. Charity$^3$, and W. H. Dickhoff$^2$} 
\affiliation{${}^1$Theory Group,
TRIUMF, Vancouver, BC V6T 2A3, Canada}
\affiliation{${}^2$Department of Physics,
Washington University, St. Louis, Missouri 63130}
\affiliation{${}^3$Department of Chemistry,
Washington University, St. Louis, Missouri 63130}

\date{\today}

\begin{abstract}
   A nonlocal dispersive-optical-model analysis has been carried out for neutrons
   and protons in $^{208}$Pb. Elastic-scattering angular distributions, total and
   reaction cross sections, single-particle energies, the neutron and proton
   numbers, the charge distribution, and the binding energy have been fitted to extract the neutron
   and proton self-energies both above and below the Fermi energy. From the
   single-particle propagator derived from these self-energies, we have
   determined the charge and matter distributions in $^{208}$Pb. 
   The predicted spectroscopic factors are consistent with results from the $(e,e'p)$ reaction and inelastic-electron-scattering data to very high spin states.
   Sensible results for the high-momentum content of neutrons and protons are obtained with protons appearing more correlated, in agreement with experiment and \textit{ab initio} calculations of asymmetric matter.
   A neutron skin of $0.25\pm0.05$~fm is deduced. An analysis of several nuclei leads
   to the conclusion that finite-size effects play a non-negligible role in the
   formation of the neutron skin in finite nuclei.
\end{abstract}

\maketitle

\section{Introduction}
The description of the properties of heavy nuclei is at present restricted to mean-field approaches.
For a nucleus like ${}^{208}$Pb, a large amount of data exists that is completely outside the scope of these methods.
In particular, elastic-nucleon-scattering data cannot be adequately accounted for with a real mean-field potential as it does not account for inelastic processes that remove flux from the elastic channel.
Properties of the ground state such as the charge density can be directly probed through elastic electron
scattering~\cite{Frois77,deVries:1987}.
Mean-field methods do not account for all the details of the deduced proton distribution, in particular in the interior of the nucleus, and are only fitted to the experimental root-mean-squared radius (rms).
Of related interest is the single-particle structure in the ground state of ${}^{208}$Pb most delicately probed with the $(e,e'p)$ reaction~\cite{Quint86,Quint87,Quint88}.
Another insight is provided by inelastic electron scattering to very high spin states~\cite{Lichtenstadt79} which was interpreted, based on results from \textit{ab initio} calculations of nuclear matter, in terms of partial occupation of single-particle orbits in ${}^{208}$Pb~\cite{Vijay84}.
Short-range properties of nuclei~\cite{Hen:2017}, as demonstrated by high-momentum components of nucleons in the ground state and their isospin dependence~\cite{Duer:2018}, provide complementary information on the ground state.
Their presence documents that mean-field orbits are depleted and need to be compensated by the occupation of nucleon states that are empty in the mean-field picture~\cite{Dickhoff04}.

A framework to encompass both ground-state properties and elastic nucleon-scattering data is provided by the dispersive optical model (DOM) originally developed by Mahaux and Sartor~\cite{Mahaux:91} and more recently reviewed in Refs.~\cite{Dickhoff:2017,Dickhoff:2019}.
The underlying formal framework of this approach is provided by the Green's function formulation of the many-body problem in which the nucleon propagator receives both particle and hole contributions, thereby inextricably linking these domains~\cite{Exposed!}.
The usual local implementation of the DOM~\cite{Mahaux:91} was extended to include fully nonlocal potentials in Ref.~\cite{Mahzoon:2014} with a complete analysis of all available ${}^{40}$Ca data including the charge density.
The subsequent results of the particle spectral density in Ref.~\cite{Dussan:2014} demonstrated that the constraint of elastic-scattering data directly provides information on the depletion of orbits which are mostly occupied in the ground state confirming the relevance of the method to quantify single-particle properties.
This was conclusively confirmed in Ref.~\cite{Atkinson:2018} where the DOM ingredients both pertaining to the overlap functions and the distorted waves provided an accurate description of  ${}^{40}$Ca$(e,e'p)^{39}$K cross sections in the relevant kinematic domain.
The latter results increased the canonical values of proton spectroscopic factors for double closed-shell nuclei~\cite{Lapikas93} by about 0.05 due to the use of nonlocal potentials to describe the proton distorted waves. 
The coincidence cross sections of the valence transitions in the ${}^{48}$Ca$(e,e'p)^{47}$K reaction are also accurately described, provided proper care is taken of the proton reaction cross sections in the DOM analysis~\cite{Atkinson:2019}.
The resulting $N-Z$ trend of the spectroscopic strength near the Fermi energy demonstrates an increased reduction of the proton removal strength with a slope that is not as large as in Ref.~\cite{Gade:2014} but larger than obtained for transfer reactions~\cite{Dickhoff:2019} and in $(p,2p)$ reactions~\cite{Aumann18,Kawase:2018}. 

While addressing all features of single-particle properties of ${}^{208}$Pb in the present work, special emphasis will be placed on the neutron distribution in the ground state.
A critical question was addressed in Ref.~\cite{Mahzoon:2017} where it was shown that when sufficient data are available for neutron scattering, in particular total cross sections, it is possible to deduce sensible predictions for the neutron distribution of ${}^{48 }$Ca employing a nonlocal DOM analysis.
The neutron distribution of nuclei is only vaguely understood.  
In particular, for a nucleus which has a large excess of neutrons
over protons, are the extra neutrons distributed evenly
over the nuclear volume or is this excess localized in the periphery of the
nucleus?  A quantitative measure is provided by the neutron
skin, defined as the difference between neutron and proton rms
radii, 
\begin{equation}
   \Delta r_{np} = r_{n} - r_{p},
   \label{eq:skin}
\end{equation}
where 
\begin{equation}
   r_{n,p}^2 = \frac{1}{N_{n,p}}\int_0^\infty dr r^4 \rho_{n,p}(r),
   \label{eq:rms}
\end{equation}
and $N_{n,p}$ is the normalization of the particle point-distributions $\rho_{n,p}(r)$. Note that the standard convention is to define the neutron skin with respect to the nucleon point-distributions, thus the size of the nucleons are not taken into account in theoretical calculations (the size of the nucleons are also factored out from experimental form factors~\cite{PREX12}). 
Accurate knowledge of the distribution of neutrons in nuclei is important for calculations of the nuclear matrix elements relevant to $\beta$-decay processes~\cite{Pastore:2018,Hyvarinen:2015}.
Furthermore, the nuclear symmetry energy, which characterizes the variation of the binding energy as a function of neutron-proton asymmetry, opposes the creation of nuclear matter with excesses of either type of nucleon. 
The extent of the neutron skin is determined by the relative strengths of the symmetry energy between the central near-saturation and peripheral less-dense regions.  Therefore, $\Delta r_{np}$ is a measure of
the density dependence of the symmetry energy around saturation~\cite{Typel01,Furnstahl02,Steiner05,RocaMaza11}. This dependence is  very important for determining many nuclear properties, including masses,
radii, fission properties, and the location of the drip lines in the chart of nuclides. Its importance extends to astrophysics for understanding supernovae and neutron stars~\cite{Horowitz01,Steiner10}, and to heavy-ion
reactions~\cite{li08}. 

Given the importance of the neutron skin in these various areas of research, a large number of studies (both  experimental and theoretical) have been devoted to it~\cite{Tsang12}. While the value of
$r_p$ can be determined quite accurately from electron scattering~\cite{Angeli:2013}, the experimental determinations of $r_n$  are typically model dependent~\cite{Tsang12}. However, the use of
parity-violating electron scattering does allow for a nearly model-independent extraction of this quantity~\cite{Horowitz98}. The present value for $^{208}$Pb extracted with this method from the PREX
collaboration at Jefferson Lab yields a skin thickness of  $\Delta r_{np}$=0.33$^{+0.16}_{-0.18}$~fm \cite{PREX12}. 
The present DOM analysis of $^{208}$Pb leads to a connection with current experimental data on the neutron skin. Unfortunately, the uncertainty from PREX is too large to constrain the majority of the theoretical
predictions of the neutron skin from mean-field calculations~\cite{Piekarewicz:2012}. Another measurement of the neutron weak form factor of $^{208}$Pb was conducted in the summer of 2019 at Jefferson Lab under
the title of PREX2. This is an updated version of the original PREX experiment which is intended to provide a much narrower error bar for the neutron skin in $^{208}$Pb.  Thus, it is timely to make a prediction
of the neutron skin now. Our analysis of $^{208}$Pb is similar to that of our previous work on $^{48}$Ca in Ref.~\cite{Mahzoon:2017}, reporting a neutron skin of $\Delta r_{np} = 0.249\pm0.023$~fm in $^{48}$Ca.
A detailed comparison of the neutrons skins of $^{208}$Pb and $^{48}$Ca will be presented in this article.


In Sec.~\ref{sec:theory} a summary of the relevant theory is presented by providing concepts of the Green's function method in Sec.~\ref{sec:prop} and the DOM in Sec.~\ref{sec:DOM}.
The result of the nonlocal DOM analysis are provided in Sec.~\ref{sec:fit}.
The neutron skin discussion is given in Sec.~\ref{sec:skin} with conclusions in Sec.~\ref{sec:conclusions}.

\section{Theory}
\label{sec:theory}

This section is organized to provide brief introductions into the underlying
theory of the method used.

   \subsection{Single-particle propagator}
\label{sec:prop}
\begin{figure*}[tb]
    \begin{center}
       \includegraphics[scale=1.0]{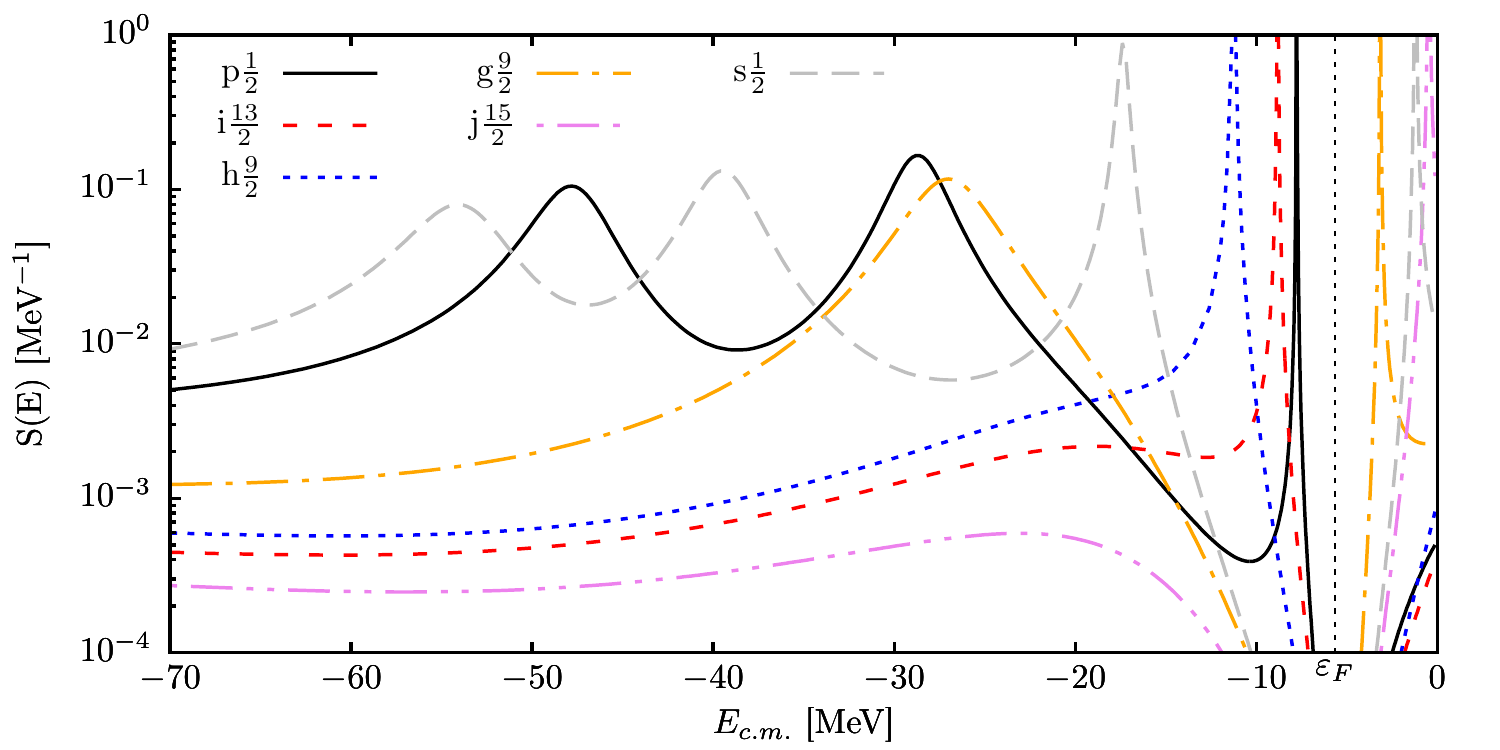}
    \end{center}
    \caption{Neutron spectral functions of a representative set of $\ell j$ shells in $^{208}$Pb. The particle states are differentiated from the hole states by the dotted line representing the Fermi energy.}
 \label{fig:spectral_n}
 \end{figure*}

\begin{figure*}[tb]
    \begin{center}
       \includegraphics[scale=1.0]{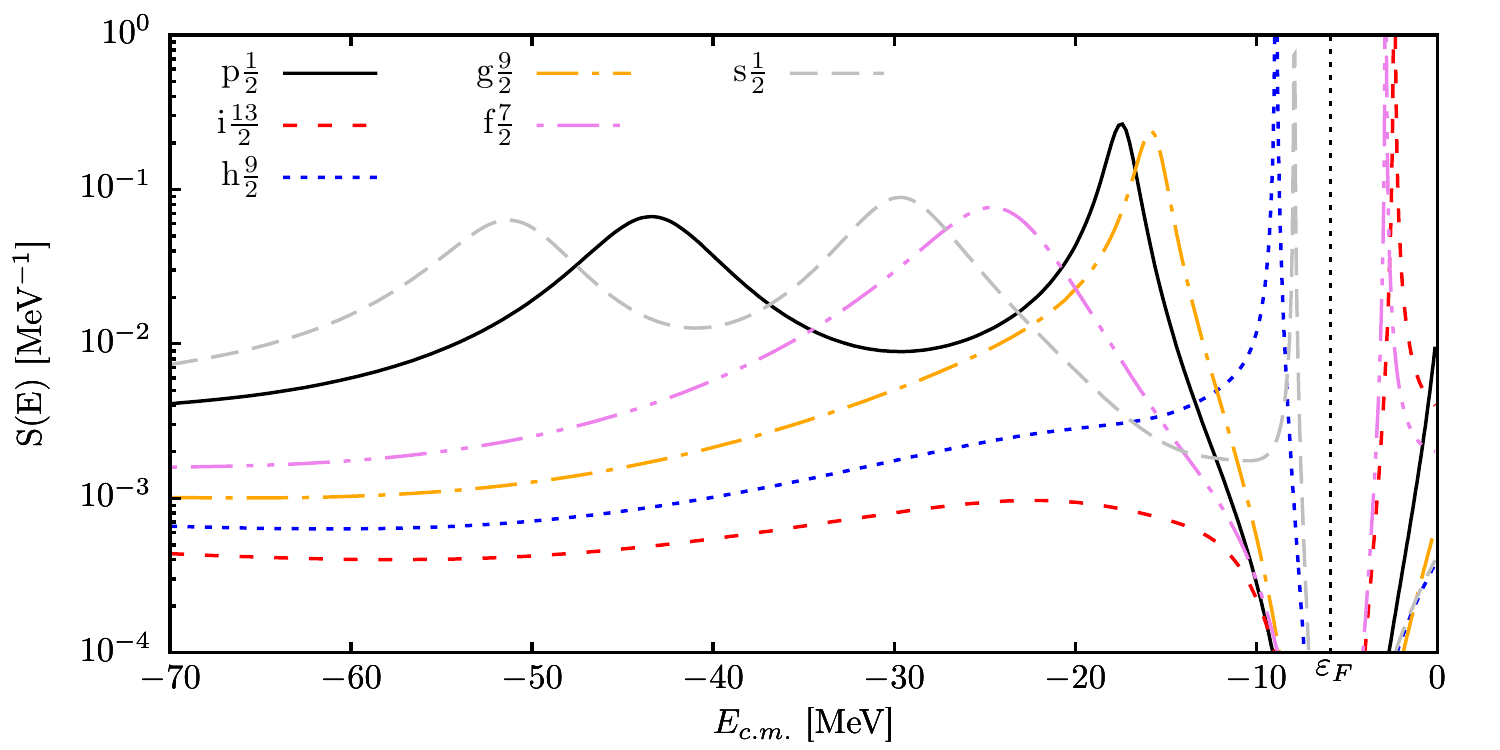}
    \end{center}
    \caption{Proton spectral functions for a representative set of $\ell j$ shells in $^{208}$Pb. The particle states are differentiated from the hole states by the dotted line representing the Fermi energy.}
 \label{fig:spectral_p}
 \end{figure*}

   The single-particle propagator describes the probability amplitude for adding (removing) a particle in state $\alpha$ at one time to the ground state and propagating on top of that state until a later time at which it is removed (added) in state $\beta$~\cite{Exposed!}.  In addition to the conserved orbital and
  total angular momentum ($\ell$ and $j$, respectively), the labels $\alpha$ and
  $\beta$ in Eq.~\eqref{eq:green} refer to a suitably chosen single-particle basis. 
 In this work the Lagrange basis~\cite{Baye:2010} was employed.
 It is convenient to work with the Fourier-transformed propagator in the energy domain, 
 \begin{align}
    G_{\ell j}(\alpha,\beta;E)  &= \bra{\Psi_0^A}a_{\alpha \ell j}
    \frac{1}{E-(\hat{H}-E_0^A)+i\eta} a_{\beta \ell j}^\dagger\ket{\Psi_0^A}
    \nonumber \\ 
    + \bra{\Psi_0^A}&a_{\beta \ell j}^\dagger\frac{1}{E-(E_0^A-\hat{H})-i\eta}
    a_{\alpha \ell j}\ket{\Psi_0^A},
    \label{eq:green}
 \end{align}
 with $E^A_0$ representing the energy of the nondegenerate ground state $\ket{\Psi^A_0}$.
 Many interactions can occur between the addition and removal of the particle (or \textit{vice versa}), all of which need to be considered to calculate the propagator. 
 No assumptions about the detailed form of the Hamiltonian $\hat{H}$ need to be made for the present discussion, but it will be assumed that a meaningful Hamiltonian exists that contains two-body and three-body contributions.
 Application of perturbation theory then leads to the Dyson equation~\cite{Exposed!} given by
 \begin{align}
    G_{\ell j}(\alpha,\beta;E) &= G_{\ell}^{(0)}(\alpha,\beta;E) \nonumber \\ +
    \sum_{\gamma,\delta}&G_{\ell}^{(0)}(\alpha,\gamma;E)\Sigma_{\ell
    j}^*(\gamma,\delta;E)G_{\ell j}(\delta,\beta;E) ,
    \label{eq:dyson}
 \end{align}
 where $G^{(0)}_{\ell}(\alpha,\beta;E)$ corresponds to the free propagator (which only includes a kinetic contribution)
 and $\Sigma_{\ell j}^*(\gamma,\delta;E)$ is the irreducible self-energy~\cite{Exposed!}. 
 The hole spectral density for energies below $\varepsilon_F$ is obtained from 
 \begin{equation}
    S^h_{\ell j}(\alpha,\beta;E) = \frac{1}{\pi}\textrm{Im}\ G_{\ell j}(\alpha,\beta;E) .
    \label{eq:spec}
 \end{equation}
 The diagonal element of Eq.~\eqref{eq:spec} is known as the (hole) spectral function identifying the probability density for the removal of a single-particle state with quantum numbers $\alpha \ell j$ at
 energy $E$. The single-particle density distribution can be calculated from the hole spectral function in the following way, 
 \begin{equation}
    \label{eq:charge}
    \rho_{\ell j}(r) = \sum_{\ell j} (2j+1) \int_{-\infty}^{\varepsilon_F}dE\ S_{\ell j}(r,r;E).
 \end{equation}
 The spectral strength for a given $\ell j$ combination can be found by summing (integrating) the spectral function according to
 \begin{equation}
    S_{\ell j}(E) = \sum_\alpha S_{\ell j}(\alpha,\alpha;E) .
    \label{eq:strength}
 \end{equation}
 The spectral strength $S_{\ell j}(E)$ is the contribution at energy $E$ to the occupation from all orbitals with $\ell j$. It reveals that the strength for a shell can be fragmented, rather than concentrated
 at the independent-particle model (IPM) energy levels.  Figure~\ref{fig:spectral_n} shows the spectral strength for a representative set of neutron shells in $^{208}$Pb that would be considered bound in the IPM. The peaks in
 Fig.~\ref{fig:spectral_n} correspond to the binding energy of the appropriate IPM orbital. For example, the s$\frac{1}{2}$ spectral function in Fig.~\ref{fig:spectral_n} has four peaks, three below
 $\varepsilon_F$ corresponding to the 0s$\frac{1}{2}$, 1s$\frac{1}{2}$, and 2s$\frac{1}{2}$ quasihole states, and one above $\varepsilon_F$ corresponding to the 3s$\frac{1}{2}$ quasiparticle state.  Comparing
 the neutron spectral functions in Fig.~\ref{fig:spectral_n} with the proton spectral functions in Fig.~\ref{fig:spectral_p} reveals that the proton peaks are broader than those of the neutrons. The broadening
 of these peaks is a consequence of the protons being more correlated than the neutrons as determined by the fit to all relevant experimental data. 

 The occupation of specific orbitals characterized by $n$ with wave functions that
 are normalized can be obtained from Eq.~\eqref{eq:spec} by folding in the corresponding wave functions~\cite{Dussan:2014},
 \begin{equation}
    S^{n-}_{\ell j}(E) = \sum_{\alpha,\beta}[\phi^n_{\ell j}(\alpha)]^*S^h_{\ell j}(\alpha,\beta;E)\phi^n_{\ell j}(\beta) .
    \label{eq:qh_strength}
 \end{equation}
 Note that this representation of the spectral strength involves off-diagonal elements of the propagator.
 The wavefunctions used in Eq.~\eqref{eq:qh_strength} are the solutions of the Dyson equation that correspond to discrete bound states with one proton removed.
 Such quasihole wave functions can be obtained from the nonlocal Schr\"{o}dinger-like equation disregarding the imaginary part
 \begin{align}
    \sum_\gamma\bra{\alpha}T_{\ell} + \textrm{Re}\ \Sigma^*_{\ell j}(\varepsilon_n^-)\ket{\gamma}\psi_{\ell j}^n(\gamma) = \varepsilon_n^-\psi_{\ell j}^n(\alpha),
    \label{eq:schrodinger}
 \end{align}
 where $\bra{\alpha}T_\ell\ket{\gamma}$ is the kinetic-energy matrix element, including the centrifugal term.
 These wave functions correspond to overlap functions
 \begin{equation}
    \psi^n_{\ell j}(\alpha) = \bra{\Psi_n^{A-1}}a_{\alpha \ell j}\ket{\Psi_0^A}, \qquad \varepsilon_n^- = E_0^A - E_n^{A-1}.
    \label{eq:wavefunction}
 \end{equation}
 Such discrete solutions to Eq.~\eqref{eq:wavefunction} exist near the Fermi energy where there is no imaginary part of the self-energy. 
 The normalization for these wave functions is the spectroscopic factor, which is given by~\cite{Exposed!}
 \begin{equation}
    \mathcal{Z}^n_{\ell j} = \bigg(1 - \frac{\partial\Sigma_{\ell j}^*(\alpha_{qh},\alpha_{qh};E)}{\partial E}\bigg|_{\varepsilon_n^-}\bigg)^{-1},
    \label{eq:sf}
 \end{equation}
 where $\alpha_{qh}$ corresponds to the quasihole state that solves Eq.~\eqref{eq:schrodinger}. This corresponds to the spectral strength at the quasihole energy $\varepsilon_n^-$, represented by a delta
 function. The quasihole peaks in Fig.~\ref{fig:spectral_p} get narrower as the levels approach $\varepsilon_F$, which is a consequence of the imaginary part of the irreducible self-energy decreasing when
 approaching $\varepsilon_F$. In fact, the last mostly occupied proton level in Fig.~\ref{fig:spectral_p} (2s$\frac{1}{2}$) has a spectral function that is essentially a delta function peaked at its energy level, where
 the imaginary part of the self-energy vanishes.  For these orbitals, the strength of the spectral function at the peak corresponds to the spectroscopic factor in Eq.~\eqref{eq:sf}. This factor can be probed
 using the exclusive $(e,e'p)$ reaction as discussed in Ref.~\cite{Atkinson:2018}.  Note that because of the presence of imaginary parts of the self-energy at other energies, there is also strength located
 there, thus the spectroscopic factor will be less than 1 and also less than the occupation probability.

\begin{figure}[tb]
    \begin{center}
       \includegraphics[scale=1.0]{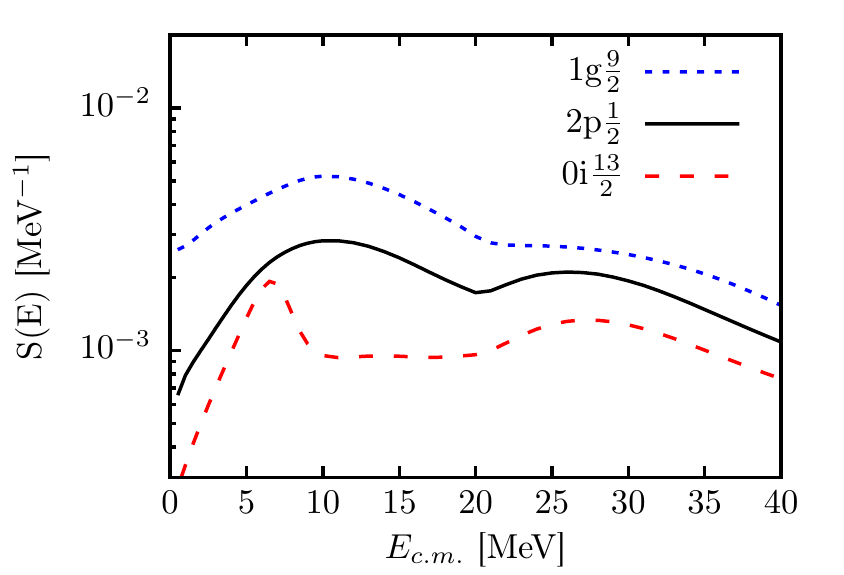}
    \end{center}
    \caption{Neutron particle spectral functions for a representative set of IPM orbitals in $^{208}$Pb that are mostly occupied but exhibit strength at positive energy which is constrained by elastic-nucleon-scattering data~\cite{Dussan:2014}.}
 \label{fig:spectral_pos}
 \end{figure}

 Indeed as shown in Ref.~\cite{Dussan:2014}, an equivalent spectral density  $S^p_{\ell j}(\alpha,\beta;E)$ for energies above $\varepsilon_F$ can be obtained which allows for the calculation of the presence of
 orbits that describe localized (and therefore normalized) single-particle states according to \begin{equation} S^{n+}_{\ell j}(E) = \sum_{\alpha,\beta}[\phi^n_{\ell j}(\alpha)]^*S^p_{\ell
    j}(\alpha,\beta;E)\phi^n_{\ell j}(\beta) .  \label{eq:qp_strength} \end{equation} Neutron spectral functions for a representative set of orbitals at positive energies are shown in
 Fig.~\ref{fig:spectral_pos}. The curve with the least strength at positive energies in Fig.~\ref{fig:spectral_pos} corresponds to the most deeply bound orbital in $^{208}$Pb. With increasing principal quantum
 number $n$, the orbital becomes less bound and the particle spectral function gains more strength at positive energies. This behavior is caused by the dispersion relation, Eq.~\eqref{eq:dispersion}, which
 pushes more strength to positive energies as the peak of the spectral function gets closer to 0 MeV. We note that the distribution at positive energies is constrained by elastic-scattering data, making the conclusion of the relevance of correlations beyond the IPM
 inevitable~\cite{Dussan:2014}. The spectral strength distribution below $\varepsilon_F$ is constrained by the charge density and particle number which also receive contributions from other $\ell j$ quantum
 numbers~\cite{Exposed!}.

 It is appropriate to introduce the Fermi energies for removal and addition given by
 \begin{equation}
    \label{eq:5.11a}
    \varepsilon^-_F = E^A_0 - E^{A-1}_0
 \end{equation}
 and
 \begin{equation}
    \varepsilon^+_F = E^{A+1}_0 - E^A_0 ,
    \label{eq:5.11b}
 \end{equation}
 referring to the ground states in the $A\pm1$ systems, respectively.
 It is also convenient to employ the average Fermi energy
 \begin{equation}
    \varepsilon_F \equiv \frac{1}{2} \left[
    \varepsilon_F^+  - \varepsilon_F^- \right]  .
    \label{eq:FE}
 \end{equation}
 In practical work, we adhere to the average Fermi energy to separate the particle and hole domain and their corresponding imaginary parts of the self-energy. 
 For specific questions related to valence holes, the imaginary part of the
 self-energy can be neglected and Eqs.~\eqref{eq:schrodinger} and \eqref{eq:sf} can be applied.
 The occupation probability of each orbital is calculated by integrating all contributions from the spectral strength up to the Fermi energy
 \begin{equation}
    n^n_{\ell j} = \int_{-\infty}^{\varepsilon_F} \!\!\!\! dE\ S^{n-}_{\ell j}(E) ,
    \label{eq:occ}
 \end{equation}
 whereas the depletion of the orbit is obtained from
 \begin{equation}
    d^n_{\ell j} = \int_{\varepsilon_F}^\infty \!\!\!\! dE\ S^{n+}_{\ell j}(E) .
    \label{eq:dep}
 \end{equation}
 Since the DOM has so far been limited to 200 MeV positive energy, a few percent of the sum rule
 \begin{equation}
    n^n_{\ell j} + d^n_{\ell j} = 1 ,
    \label{eq:antic}
 \end{equation}
 that reflects the anticommutator relation of the corresponding fermion addition and removal operators, 
 has been found above this energy~\cite{Dussan:2014}.
 The particle number of the nucleus is found by summing over each $\ell j$ combination while integrating the spectral strength up to the Fermi energy, 
 \begin{equation}
    Z,N = \sum_{\ell j} (2j+1) \int_{-\infty}^{\varepsilon_F} \!\!\!\! dE\  S^{p,n}_{\ell j}(E) ,
    \label{eq:particle}
 \end{equation}
 where $Z$ and $N$ are the total number of protons and neutrons, respectively. 
 In addition to particle number, the total binding energy can be calculated from the hole spectral function using the Migdal-Galitski sum rule~\cite{Exposed!}, 
 \begin{align}
    E_0^{N,Z} &= \frac{1}{2}\sum_{\alpha\beta}\int_0^{\varepsilon_F}dE\left[\braket{\alpha|\hat{T}|\beta}S^h(\alpha,\beta;E)\right.\\
    &\left. + \delta_{\alpha\beta}ES^h(\alpha,\alpha;E)\vphantom{\braket{\alpha|\hat{T}|\beta}S^h(\alpha,\beta;E)}\right].
    \label{eq:energy_sumrule}
 \end{align}

 \subsection{Dispersive optical model}
 \label{sec:DOM}
 It was recognized long ago that the irreducible self-energy represents the potential 
 that describes elastic-scattering observables~\cite{Bell59}. 
 The link with the potential at negative energy is then provided by the Green's function framework as was realized by Mahaux and Sartor who introduced the DOM as reviewed in Ref.~\cite{Mahaux:91}. 
 The analytic structure of the nucleon self-energy allows one 
 to apply the dispersion relation, which relates the real part of the self-energy at a given energy to a dispersion integral of its imaginary part over all energies.
 The energy-independent correlated Hartree-Fock (HF) contribution~\cite{Exposed!} is removed by employing a subtracted dispersion relation with the Fermi energy used as the subtraction point~\cite{Mahaux:91}.
 The subtracted form has the further advantage that the emphasis is placed on energies closer to the Fermi energy for which more experimental data are available.
 The real part of the self-energy at the Fermi energy is then still referred to as the HF term, but is sufficiently attractive to bind the relevant levels.
 In practice, the imaginary part is assumed to extend to the Fermi energy on both sides while being very small in its  vicinity.
 The subtracted form of the dispersion relation employed in this work is given by
 \begin{align}
    \textrm{Re}\ \Sigma^*(\alpha,\beta;E) &= \textrm{Re}\
    \Sigma^*(\alpha,\beta;\varepsilon_F) \label{eq:dispersion} \\ -
    \mathcal{P}\int_{\varepsilon_F}^{\infty} \!\! \frac{dE'}{\pi}&\textrm{Im}\
    \Sigma^*(\alpha,\beta;E')[\frac{1}{E-E'}-\frac{1}{\varepsilon_F-E'}] \nonumber
    \\ + \mathcal{P} \! \int_{-\infty}^{\varepsilon_F} \!\!
    \frac{dE'}{\pi}&\textrm{Im}\
    \Sigma^*(\alpha,\beta;E')[\frac{1}{E-E'}-\frac{1}{\varepsilon_F-E'}],
    \nonumber      
 \end{align}
 where $\mathcal{P}$ is the principal value. 
 The static term is denoted by  $\Sigma_{\text{HF}}$ from here on. 
 Equation~\eqref{eq:dispersion} constrains the real part of the self-energy through empirical information of the HF term and empirical knowledge of the imaginary part, which is closely tied to experimental data. 
 Initially, standard functional forms for these terms were introduced by Mahaux and Sartor who also cast the DOM potential in a local form by a standard transformation which turns a nonlocal static HF potential into an energy-dependent local potential~\cite{Perey:1962}.
 Such an analysis was extended in Refs.~\cite{Charity06,Charity:2007} to a sequence of Ca isotopes and in Ref.~\cite{Mueller:2011} to semi-closed-shell nuclei heavier than Ca.
 The transformation to the exclusive use of local potentials precludes a proper calculation of nucleon particle number and expectation values of the one-body operators, like the charge density in the ground state. 
 This obstacle was eliminated in Ref.~\cite{Dickhoff:2010}, but it was shown that the introduction of nonlocality in the imaginary part was still necessary in order to accurately account for particle number and the charge density~\cite{Mahzoon:2014}.
 Theoretical work provided further support for this introduction of a nonlocal representation of the imaginary part of the self-energy~\cite{Waldecker:2011,Dussan:2011}.
 A recent review has been published in Ref.~\cite{Dickhoff:2017}.

 We implement a nonlocal representation of the self-energy following
 Ref.~\cite{Mahzoon:2014} where $\Sigma_{\text{HF}}(\bm{r},\bm{r'})$ and
 $\textrm{Im}\ \Sigma(\bm{r},\bm{r'};E)$ are parametrized, using Eq.~\eqref{eq:dispersion} to generate the energy dependence of the
 real part. The HF term consists of a volume term, spin-orbit term,  and a wine-bottle-shaped term~\cite{Brida11} to simulate a surface contribution. The imaginary self-energy consists of volume, surface, and spin-orbit terms. 
 Details can be found in App.~\ref{Sec:param}. Nonlocality is represented using the Gaussian form 
 \begin{equation}
    H(\bm{s},\beta) = \pi^{-3/2}\beta^{-3}e^{-\bm{s}^2/\beta^2} ,
    \label{eq:nonlocality}
 \end{equation}
 where $\bm{s} = \bm{r} -\bm{r}'$, 
 as proposed in Ref.~\cite{Perey:1962}. 
 As mentioned previously, it was customary in the past to replace nonlocal potentials by local, energy-dependent potentials~\cite{Mahaux:91,Perey:1962,Fiedeldey:1966,Exposed!}. The introduction of an energy dependence alters the dispersive
 correction from Eq.~\eqref{eq:dispersion} and distorts the normalization, leading to incorrect spectral functions and related quantities~\cite{Dickhoff:2010}. Thus, a nonlocal implementation permits the self-energy to accurately 
 reproduce important observables such as the charge density and particle number. 

 In order to use the DOM self-energy for predictions, the parameters are fit through a weighted $\chi^2$ minimization of available elastic differential cross section data ($\frac{d\sigma}{d\Omega}$), analyzing power data ($A_\theta$),  reaction cross sections ($\sigma_r$), total cross sections ($\sigma_t$), charge density ($\rho_{\text{ch}}$), energy levels ($\varepsilon_{n \ell j}$), particle number, separation energies, 
 and the root-mean-square charge radius ($r_{\text{rms}}$).
 While it has been suggested in Refs.~\cite{Pawel17,Loc:2014,Khoa:2007} that $(p,n)$ cross sections to
isobaric analogue states provide additional information on the isovector
potential, our current implementation of the DOM does not include these data.
We checked that reasonable cross sections are obtained with our DOM potential,
suggesting that these data, while important, are not sufficient to alter the
conclusions of our work significantly. This may be due to the use of nonlocal
potentials as opposed to the local ones used in Refs.~\cite{Loc:2014,Khoa:2007} based on Ref.~\cite{Koning:2003}. We
plan in future applications to include these data for additional nuclei in a
more consistent manner. 

 The potential is transformed from coordinate-space to a Lagrange basis using Legendre and Laguerre polynomials for scattering and bound states, respectively.
 The bound states are found by diagonalizing the Hamiltonian in
 Eq.~\eqref{eq:schrodinger}, the propagator is found by inverting the Dyson
 equation, Eq.~\eqref{eq:dyson}, 
 while all scattering calculations are done in the framework of $R$-matrix theory~\cite{Baye:2010}. 
 Implementations of the nonlocal DOM in $^{40}$Ca and $^{48}$Ca have previously been published in Refs.~\cite{Mahzoon:2017,Atkinson:2018,Mahzoon:2014}.

 \section{DOM fit of $^{208}$Pb}
 \label{sec:fit}

 The functional form of the $^{208}$Pb self-energy is equivalent to that of
 $^{48}$Ca used in Ref.~\cite{Mahzoon:2017}. Starting from the parameters for $^{48}$Ca, the $\chi^2$ was minimized
 for a similar set of experimental data for $^{208}$Pb (see App.~\ref{app:params} for
 specific values of parameters). 
In the analysis presented here, minimization was performed using an implementation of the Powell method~\cite{Numerical}. Due to computational challenges of parameter fitting with this method and to cross-validate our approach, we also conducted a parallel DOM analysis of $^{208}$Pb using Markov Chain Monte Carlo (MCMC) to optimize the potential parameters, using the same experimental data and a very similar functional form for the self-energy. The preliminary spectroscopic factor, neutron skin, and spectral function results of this parallel analysis are in excellent agreement (\textit{e.g.}, all within one standard deviation) with those detailed in the following sections and will be the subject of a subsequent publication by our group.

\begin{figure}[tb]
    \begin{minipage}{\linewidth}
       \makebox[\linewidth]{
          \includegraphics[scale=1.0]{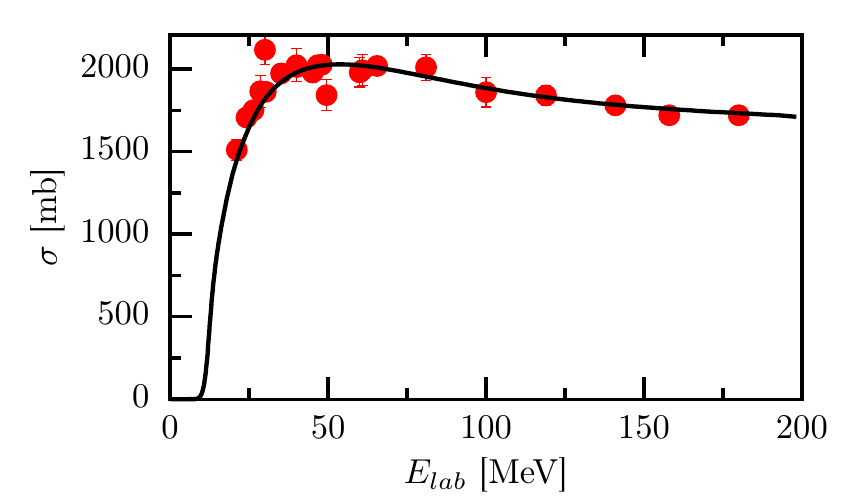}
       }
    \end{minipage}
    \caption[The proton reaction cross section for $^{208}$Pb.]{The proton reaction
       cross section for $^{208}$Pb. References to the experimental data points can
       be found in Ref.~\cite{Mueller:2011}.
    }
    \label{fig:react_pb208}
 \end{figure}

\begin{figure}[tb]
    \begin{minipage}{\linewidth}
       \makebox[\linewidth]{
          \includegraphics[scale=1.0]{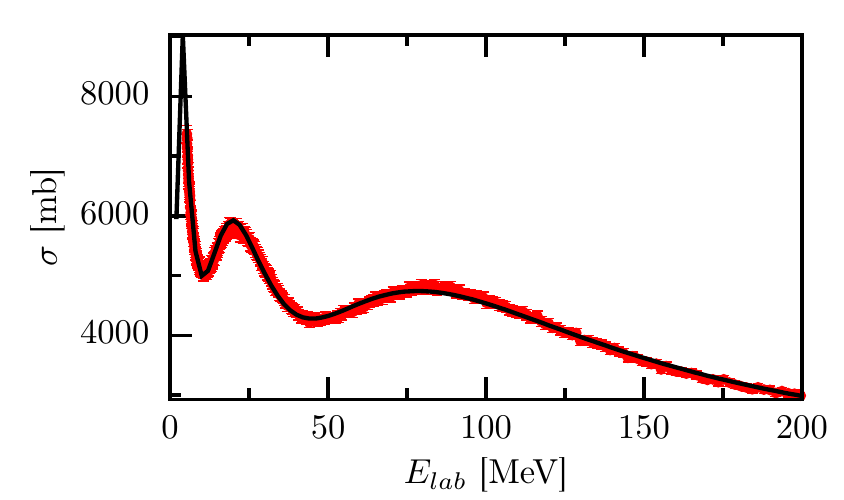}
       }
    \end{minipage}
    \caption[Neutron total cross section of $^{208}$Pb]{Neutron total cross section
       (solid line) generated from the DOM
       self-energy for $^{208}$Pb. The circles represent measured total cross sections. References to the data are given in Ref.~\cite{Mueller:2011}.
    }
    \label{fig:total_pb208}
 \end{figure}

\begin{figure}[tb]
    \begin{minipage}{\linewidth}
       \makebox[\linewidth]{
          \includegraphics[scale=1.0]{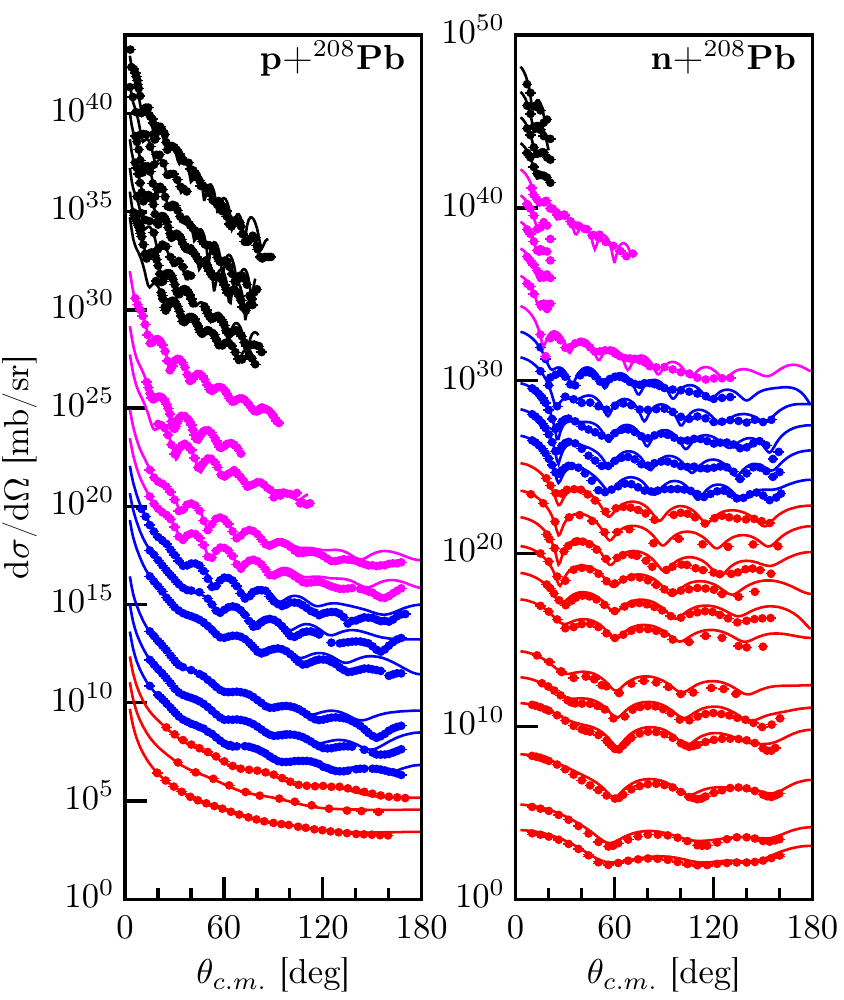}
       }
    \end{minipage}
    \caption[Calculated and experimental proton and neutron elastic-scattering
       angular distributions of the differential cross section
    $\frac{d\sigma}{d\Omega}$ for $^{208}$Pb]{Calculated and experimental proton and neutron
       elastic-scattering angular distributions of the differential cross section
       $\frac{d\sigma}{d\Omega}$ for $^{208}$Pb ranging from 10 MeV - 200 MeV. The data at each energy is offset by factors of
       ten to help visualize all of the data at once. References to the data are
       given in Ref.~\cite{Mueller:2011}. 
    }
    \label{fig:elastic_pb208}
 \end{figure}

\begin{figure}[tb]
    \begin{minipage}{\linewidth}
       \makebox[\linewidth]{
          \includegraphics[scale=1.0]{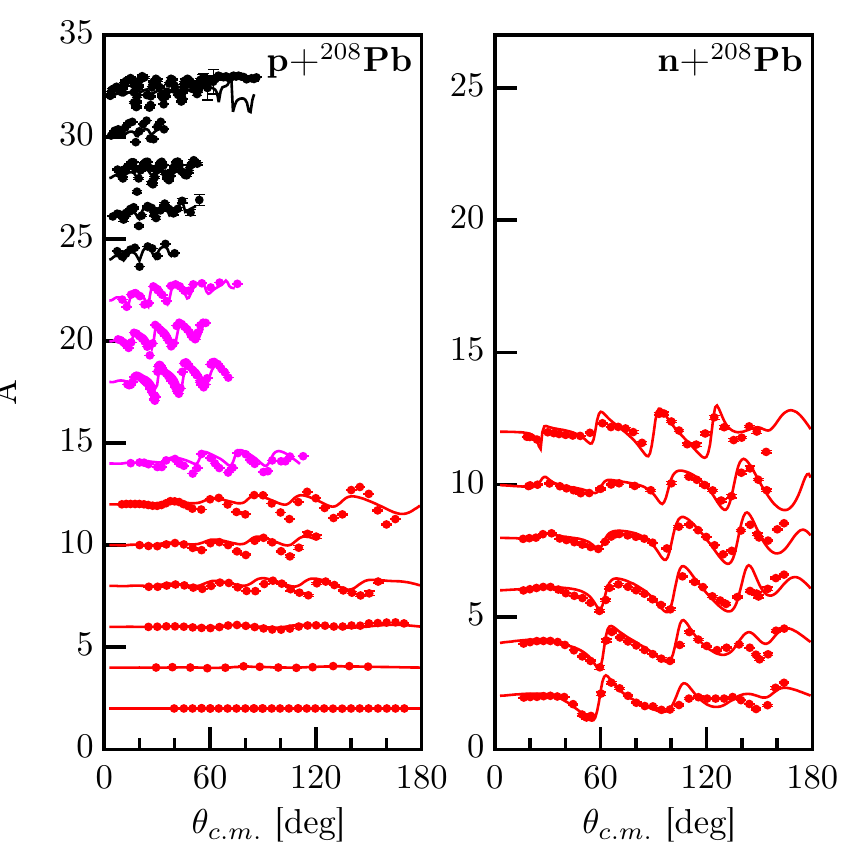}
       }
    \end{minipage}
    \caption[Proton and neutron analyzing power generated from the DOM
    self-energy for $^{208}$Pb compared with experimental data]{Results for proton
       and neutron analyzing power generated from the DOM self-energy for
       $^{208}$Pb compared with experimental data ranging from 10 MeV - 200 MeV. References to the data are given
       in Ref.~\cite{Mueller:2011}.
    }
    \label{fig:analyze_pb208}
 \end{figure}

 Proton reaction cross sections together with the DOM result are displayed in Fig.~\ref{fig:react_pb208}.
 The neutron total cross sections
 are shown in Fig.~\ref{fig:total_pb208}.  
 Both aggregate cross sections play an important role in determining volume integrals of the imaginary part of the self-energy, thereby providing strong constraints on the depletion of IPM orbits.
 The elastic
 differential cross sections at energies up to 200 MeV
 for protons and neutrons are shown in Fig.~\ref{fig:elastic_pb208}. The
 analyzing powers for neutrons and protons are shown in
 Fig.~\ref{fig:analyze_pb208}.  

 \begin{figure}[tb]
    \begin{minipage}{\linewidth}
       \makebox[\linewidth]{
          \includegraphics[scale=1.0]{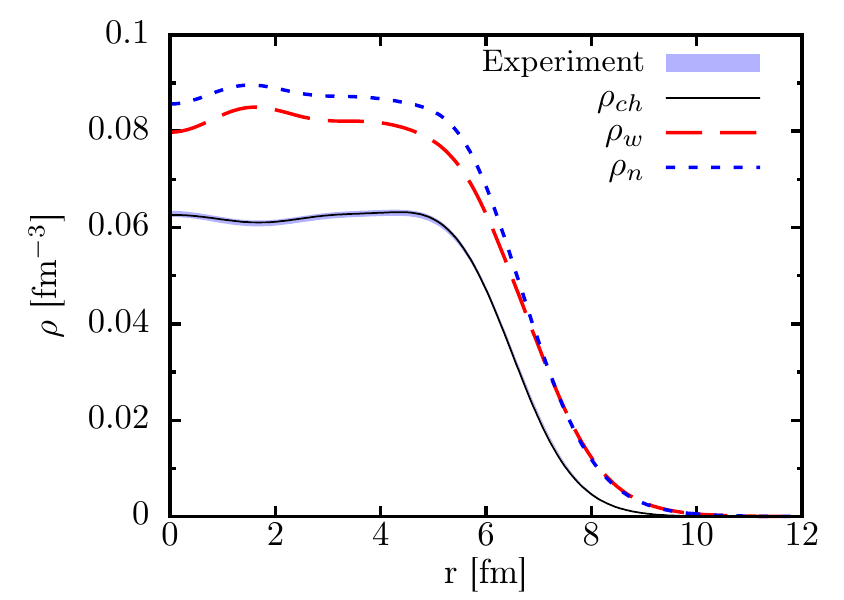}
       }
    \end{minipage}
    \caption[Experimental and fitted $^{208}$Pb charge density.]{Experimental
       and fitted $^{208}$Pb charge density. The solid black line is calculated
       using Eq.~\eqref{eq:charge} and folding with the proton
       charge distribution while the experimental band represents the 1\%
       error associated with the extracted charge density from elastic
       electron scattering experiments using the sum of Gaussians
       parametrization~\cite{deVries:1987,Sick79}. Also shown is the deduced weak charge distribution, $\rho_w$ (red long-dashed line), and neutron matter distribution, $\rho_n$ (blue short-dashed line).
    }
    \label{fig:chd_pb208}
 \end{figure} 

\begin{figure}[tb]
    \begin{minipage}{\linewidth}
       \makebox[\linewidth]{
          \includegraphics[scale=1.0]{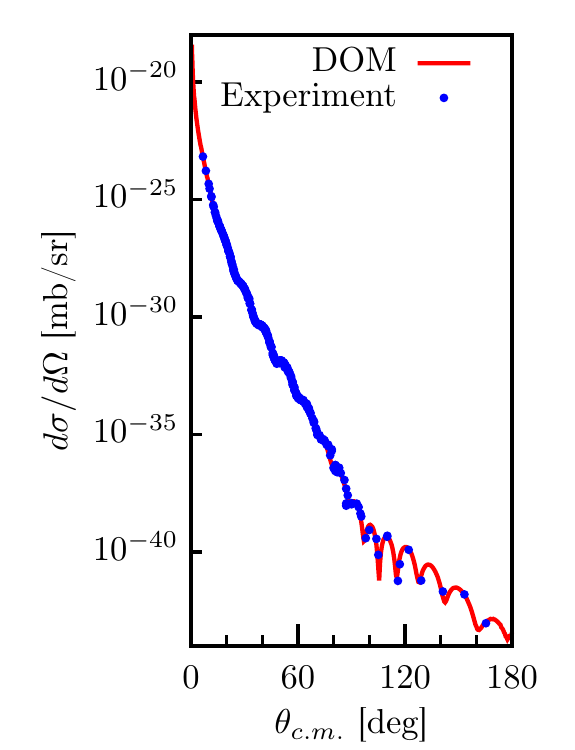}
       }
    \end{minipage}
    \caption{Experimental and fitted $^{208}$Pb$(e,e)$ differential cross section. All available data have been transformed to an electron energy of 502 MeV~\cite{Frois77}.}
 \label{fig:pb208_ee}
\end{figure}

 The charge density of $^{208}$Pb is shown in Fig.~\ref{fig:chd_pb208}. The
 experimental band is extrapolated from elastic electron scattering differential
 cross sections~\cite{deVries:1987}. This data is well reproduced after using the
 DOM charge density from Fig.~\ref{fig:chd_pb208} as the ingredient in a
 relativistic elastic electron scattering code~\cite{salvat:2005}. The corresponding elastic electron 
 scattering cross section is shown in Fig.~\ref{fig:pb208_ee} and compared to experiment with all available data transformed to an electron energy of 502 MeV~\cite{Frois77}.

\begin{figure}[tb]
   \begin{minipage}{\linewidth}
      \makebox[\linewidth]{
         \includegraphics[scale=1.0]{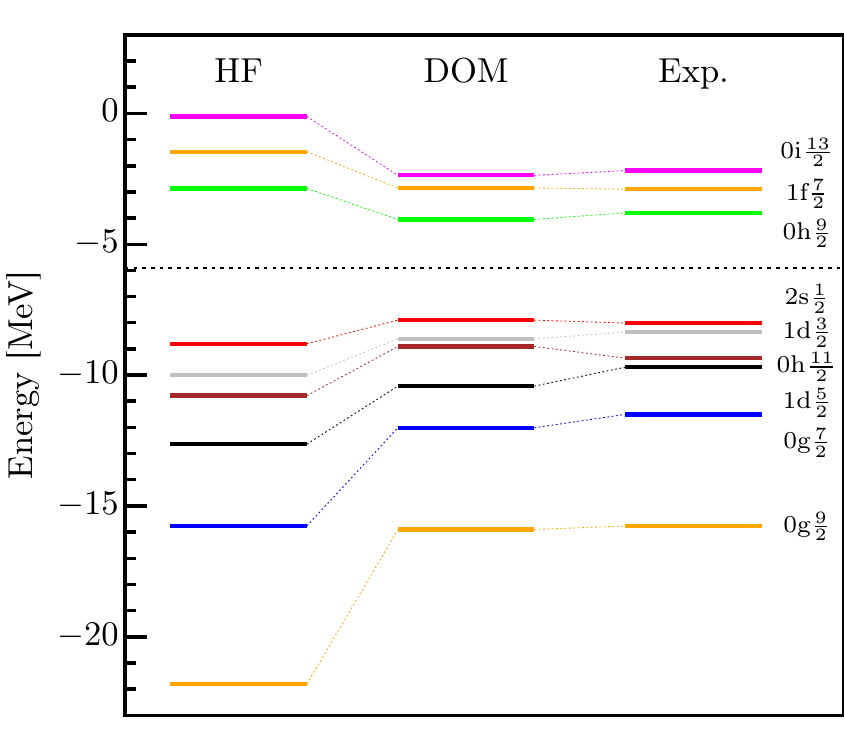}
      }
   \end{minipage}
   \caption{Proton energy levels in $^{208}$Pb. The energies on the left are calculated using only the static part of the DOM self-energy, corresponding to a Hartree-Fock calculation. The middle energies are those calculated using the full DOM self-energy. The energy on the right correspond to the experimental values. The change from the left energies to the middle energies is the result of including the dynamic part of the self-energy.}
   \label{fig:levelsP}
\end{figure}

\begin{figure}[tb]
   \begin{minipage}{\linewidth}
      \makebox[\linewidth]{
         \includegraphics[scale=1.0]{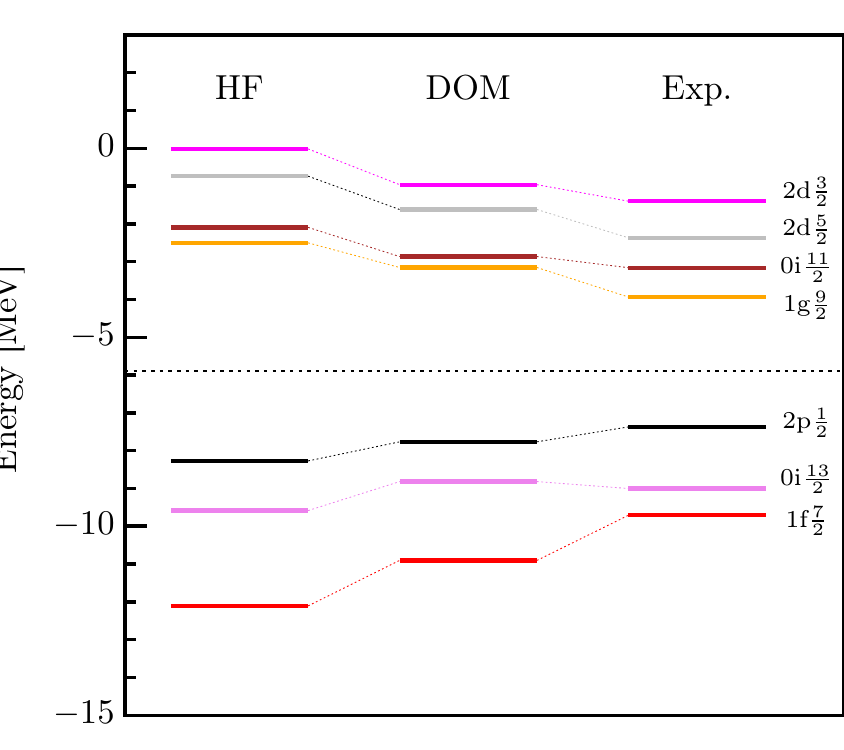}
      }
   \end{minipage}
   \caption{Neutron energy levels in $^{208}$Pb. The energies on the left are calculated using only the static part of the DOM self-energy, corresponding to a Hartree-Fock calculation. The middle energies are those calculated using the full DOM self-energy. The energy on the right correspond to the experimental values. The change from the left energies to the middle energies is the result of including the dynamic part of the self-energy.}
   \label{fig:levelsN}
\end{figure}

In Figs.~\ref{fig:levelsP} and~\ref{fig:levelsN}, single-particle levels calculated using Eq.~\eqref{eq:schrodinger} are compared to the experimental values for protons and neutrons, respectively. 
The middle column consists of levels calculated using the full DOM and the right column contains the experimental levels. The first column of the figures represents a calculation using only the static part of
the self-energy, corresponding to the Hartree-Fock (mean-field) contribution. It is clear from these level diagrams that the mean-field overestimates the particle-hole gap (see also Ref.~\cite{Bender:2003}). The inclusion of the dynamic part of
the self-energy is necessary to reduce this gap and properly describe the energy levels. Furthermore, the effect of including the dynamic part of the self-energy on the proton levels is stronger than the
effect on the neutron levels. This is another manifestation of the fact that the proton properties deviate more from the IPM than the neutrons in $^{208}$Pb.

\begin{table}[tb]
   \caption
   {DOM calculated spectroscopic factors for orbitals with energy levels near $\varepsilon_F$.}
   \label{table:sf208}
   \begin{center}
      \begin{tabular}{cc} 
         \hline
         \hline
         \textbf{Protons} & \textbf{$\mathcal{Z}_{\ell j}^n$}\\
         \hline
         $\mathbf{1f_{7/2}}$ & 0.67\\
         \hline
         $\mathbf{0h_{9/2}}$ & 0.60\\
         \hline
         $\mathbf{2s_{1/2}}$ & 0.69\\
         \hline
         $\mathbf{1d_{3/2}}$ & 0.66\\
         \hline
         $\mathbf{0h_{11/2}}$ & 0.61\\
         \hline
         $\mathbf{1d_{5/2}}$ & 0.68\\
         \hline
         \hline
      \end{tabular}
      \begin{tabular}{cc} 
         \hline
         \hline
         \textbf{Neutrons} & \textbf{$\mathcal{Z}_{\ell j}^n$}\\
         \hline
         $\mathbf{0i_{11/2}}$ & 0.77\\
         \hline
         $\mathbf{1g_{9/2}}$ & 0.77\\
         \hline
         $\mathbf{2p_{1/2}}$ & 0.81\\
         \hline
         $\mathbf{1f_{5/2}}$ & 0.81\\
         \hline
         $\mathbf{2p_{3/2}}$ & 0.82\\
         \hline
         $\mathbf{0i_{13/2}}$ & 0.80\\
         \hline
         \hline
      \end{tabular}
   \end{center}
\end{table}

For levels close to $\varepsilon_F$, the spectroscopic factor can be calculated using Eq.~\eqref{eq:sf}. These spectroscopic factors are listed in Table~\ref{table:sf208}. Indeed, the fact that the
spectroscopic factors for protons are smaller than those of the neutrons is consistent with the protons being more correlated than the neutrons. 
The present values of the valence spectroscopic factors are consistent with the observations of Ref.~\cite{Lichtenstadt79} and the interpretation of Ref.~\cite{Vijay84}.
It is important to note that these spectroscopic factors are indirectly determined by the fit to all the available data similar to the case reported in Ref.~\cite{Atkinson:2018} for ${}^{48}$Ca.
The extraction of spectroscopic factors using the $(e,e'p)$ reaction has yielded a value around 0.65 for the valence 
$\mathrm{2s_{1/2}}$ orbit~\cite{Ingo91} based on the results of Ref.~\cite{Quint86,Quint87}.
While the use of nonlocal optical potentials may slightly increase this value as shown in Ref.~\cite{Atkinson:2018}, it may be concluded that the value of 0.69 obtained from the present analysis is completely consistent with this result.
Nikhef data obtained in a large missing energy and momentum domain~\cite{Batenburg01} can therefore now be consistently analyzed employing the complete DOM spectral functions.

\begin{table}[tb]
   \caption
   {Calculated DOM occupation and depletion for the orbitals shown in Figs.~\ref{fig:levelsP} and~\ref{fig:levelsN}.}
   \label{table:occ}
   \begin{center}
      {\renewcommand{\arraystretch}{1.15}
      \begin{tabular}{ccc} 
         \hline
         \hline
         \bf{Protons} & $n_{\ell j}^n$ & $d_{\ell j}^n$\\
         \hline
         \bf{2s$\frac{1}{2}$} & 0.76 & 0.088 \\
         \hline
         \bf{1d$\frac{3}{2}$} & 0.77 & 0.015 \\
         \hline
         \bf{1d$\frac{5}{2}$} & 0.78 & 0.014 \\
         \hline
         \bf{1f$\frac{7}{2}$} & 0.051 & 0.68 \\
         \hline
         \bf{0g$\frac{7}{2}$} & 0.80 & 0.0065 \\
         \hline
         \bf{0g$\frac{9}{2}$} & 0.81 & 0.0054 \\
         \hline
         \bf{0h$\frac{9}{2}$} & 0.082 & 0.66 \\
         \hline
         \bf{0h$\frac{11}{2}$} & 0.73 & 0.0066 \\
         \hline
         \bf{0i$\frac{13}{2}$} & 0.054 & 0.75 \\
         \hline
      \end{tabular}
   }
   {\renewcommand{\arraystretch}{1.15}
   \begin{tabular}{ccc} 
      \hline
      \hline
      \bf{Neutrons} & $n_{\ell j}^n$ & $d_{\ell j}^n$\\
      \hline
      \bf{2p$\frac{1}{2}$} & 0.85 & 0.11 \\
      \hline
      \bf{2d$\frac{3}{2}$} & 0.020 & 0.96 \\
      \hline
      \bf{2d$\frac{5}{2}$} & 0.020 & 0.95 \\
      \hline
      \bf{1f$\frac{7}{2}$} & 0.88 & 0.080 \\
      \hline
      \bf{1g$\frac{9}{2}$} & 0.025 & 0.94 \\
      \hline
      \bf{0i$\frac{11}{2}$} & 0.040 & 0.92 \\
      \hline
      \bf{0i$\frac{13}{2}$} & 0.87 & 0.070 \\
      \hline
      \hline
   \end{tabular}
}
   \end{center}
\end{table}

The number of neutrons and protons in
the DOM fit of $^{208}$Pb, calculated using Eq.~\eqref{eq:particle} using shells up to $\ell \le 20$, is shown in Table~\ref{table:sumrules}.  As there are 82 protons and 126 neutrons in $^{208}$Pb, the
reported values are accurate to within a fraction of a percent.  The binding energy of $^{208}$Pb was fit to the experimental value using Eq.~\eqref{eq:energy_sumrule}. 
As there is no way at present to assess the value of three-body interactions to the ground-state energy, we employ the present approximation which applies when only two-body interactions occur in the Hamiltonian, to ensure that enough spectral strength occurs at negative energy which has implications for the presence of high-momentum components.
The comparison to the
experimental value is also shown in Table~\ref{table:sumrules}. 

\begin{figure}[tb]
   \begin{minipage}{\linewidth}
      \makebox[\linewidth]{
         \includegraphics[scale=1.0]{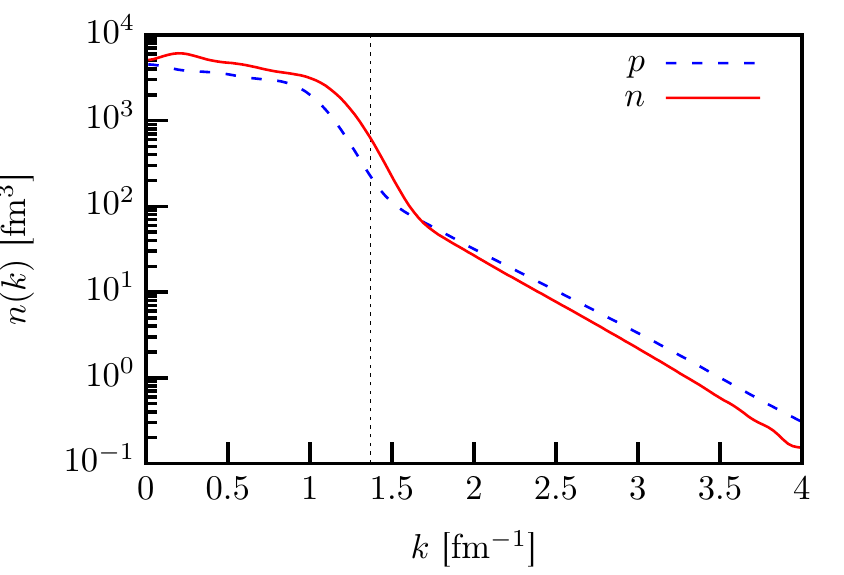}
      }
   \end{minipage}
   \caption[Comparison of calculated DOM momentum distribution of protons and
   neutrons in $^{208}$Pb.]
   {
      Comparison of calculated DOM momentum distributions of protons (blue dashed line) and neutrons (red solid line) in $^{208}$Pb. The dotted line marks the location of $k_F$. 
   }
   \label{fig:kdist}
\end{figure}

Consider the momentum distribution, $n(k)$, which is the double Fourier-transform of the single-particle density matrix, 
\begin{align}
   n(k) = \frac{2}{\pi}\sum_{\ell j}(2j+1)\int_0^\infty& dr r^2 \int_0^\infty dr' r'^2 \nonumber \\
   &\times j_\ell(kr) \rho_{\ell j}(r,r') j_\ell(kr').
   \label{eq:density_kspace}
\end{align}
The calculated DOM momentum distribution of $^{208}$Pb is shown in Fig.~\ref{fig:kdist}. The high-momentum tail of the momentum distribution arises from short-range correlations (SRC), which is another
manifestation of many-body correlations beyond the IPM description of the nucleus~\cite{Hen:2017}. This high-momentum content can be quantified by integrating the momentum distribution above the Fermi momentum.
Using $k_F=270$ MeV/c, 13.4\% of protons and 10.7\% of neutrons have momenta greater than $k_F$.  
If instead a cut-off is used of 330 MeV/c, the proton content is 8.4\%, whereas only 4.5\% neutrons are obtained.
These numbers are in qualitative agreement with what is observed in the high-momentum knockout experiments done by the CLAS
collaboration at Jefferson Lab~\cite{CLAS:2006}.  Furthermore, the fraction of high-momentum protons is larger than the fraction of high-momentum neutrons. 
These observations were predicted by \textit{ab initio} calculations of asymmetric nuclear matter reported in Refs.~\cite{Frick:2005,Rios:2009,Rios:2014} which demonstrated unambiguously that the inclusion of the nucleon-nucleon tensor force when it is constrained by nucleon-nucleon scattering data, is responsible for making protons more correlated with increasing nucleon asymmetry at normal density.
These results should come as no surprise, since
Figs.~\ref{fig:spectral_n},~\ref{fig:spectral_p},~\ref{fig:levelsP},~\ref{fig:levelsN}, and Table~\ref{table:sf208} all reveal that the protons are more correlated than the neutrons in $^{208}$Pb. This supports the $np$-dominance picture in which
the dominant contribution to SRC pairs comes from $np$ SRC pairs which arise from the tensor force in the nucleon-nucleon interaction~\cite{Duer:2018,Wiringa:2014}. Due to the neutron excess in $^{208}$Pb,
there are more neutrons available to make $np$ SRC pairs which leads to an increase in the fraction of high-momentum protons. 

In the DOM, this high-momentum content is determined by how much strength exists in the hole spectral function at large, negative energies. The hole spectral function is constrained in the fit by the particle
number, binding energy, and charge density. While the particle number and charge density can only constrain the total strength of the hole spectral function, the binding energy constrains how the strength of
the spectral function is distributed in energy. This arises from the energy-weighted integral in Eq.~\eqref{eq:energy_sumrule}, which will push some of the strength of the spectral function to more-negative
energies in order to acheive more binding. This, in turn, alters the momentum distribution, thus constraining the high-momentum content.

The reproduction of all available experimental data indicates that a suitable self-energy of $^{208}$Pb has been found. With this self-energy we can therefore make predictions of other observables, such as the neutron skin.

\begin{table}[tb]
   \caption{Comparison of the calculated DOM particle numbers and binding energy of $^{208}$Pb and the corresponding experimental values. The experimental binding energy can be found in Ref.~\cite{AME:2003}.}
   \label{table:sumrules}
   \begin{minipage}{\linewidth}
      \makebox[\linewidth]{
         {\renewcommand{\arraystretch}{1.15}
         \begin{tabular}{c  c c  c c} 
            \hline
            \hline
            & N & Z & DOM $E_0^A/A$ & Exp. $E_0^A/A$ \\
            \hline
            $^{208}$Pb & 126.2 & 82.08 & -7.82 & -7.87\\
            \hline
            \hline
         \end{tabular}
      }
   }
\end{minipage}
\end{table}

\section{Neutron Skin}
\label{sec:skin}

\begin{figure}[tb]
   \begin{minipage}{\linewidth}
      \makebox[\linewidth]{
         \includegraphics[scale=1.0]{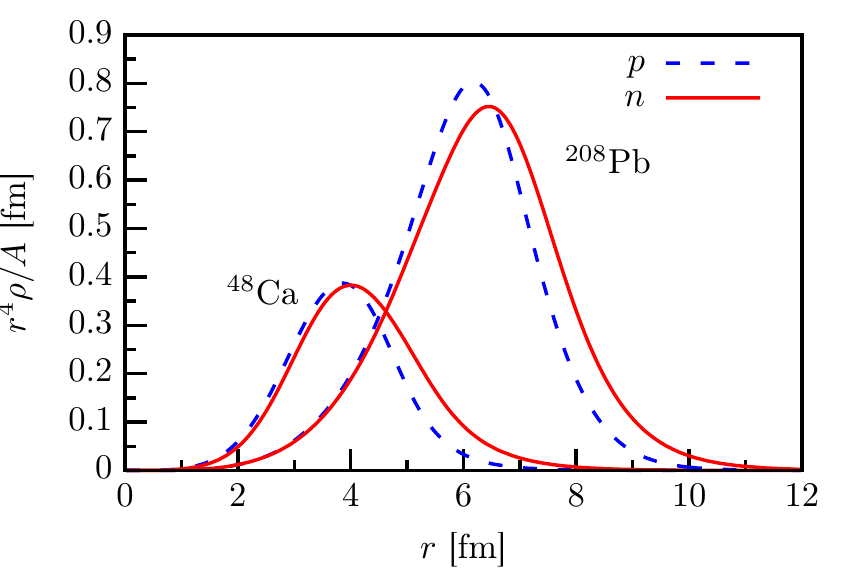}
      }
   \end{minipage}
   \caption[Neutron and proton point distributions in $^{208}$Pb weighted by $r^4$]{Neutron (red solid line)
      and proton (blue dashed line) point distributions in $^{208}$Pb and $^{48}$Ca weighted by $r^4$ and
      normalized according to Eq.~\eqref{eq:rms}.
   }
   \label{fig:dist_comp}
\end{figure} 
The neutron and proton point distributions in $^{208}$Pb, weighted by $r^4$ and normalized by particle number, are shown in Fig.~\ref{fig:dist_comp}. It is clear that the neutrons are more extended than the
protons, giving rise to a positive neutron skin of $\Delta r_{np} = 0.25\pm0.05$ fm. 
The associated error is obtained in the same manner as in Ref.~\cite{Mahzoon:2017} for ${}^{48}$Ca (in the ongoing MCMC-enabled analysis mentioned in Sec.~\ref{sec:fit}, we recover a compatible, somewhat smaller neutron skin of 0.195, with a similar uncertainty but employing a more restricted set of parameters).
It is no surprise that the value of the skin falls within the range of allowed values from the PREX experiment, but it will
be interesting to compare this prediction to the updated experimental value from PREX2 in the near future as well as new  results from the Mainz facility~\cite{Becker2018}.
This is also within the range of skin values ($0.12$ - $0.28$~fm) of the 48 nuclear energy density functionals used in
Ref.~\cite{Piekarewicz:2012}. Currently, \textit{ab initio} calculations cannot be applied to heavy systems such as $^{208}$Pb, so these mean-field results are the only other theoretical predictions of the
neutron skin in $^{208}$Pb.

The DOM predictions of the neutron skin of $^{40}$Ca, $^{48}$Ca, and $^{208}$Pb are shown in Table.~\ref{table:skins_normalized}, where it is evident that the neutron skins of $^{48}$Ca and $^{208}$Pb are very
similar. Since $^{208}$Pb and $^{48}$Ca have similar asymmetry parameters, indicated by $\alpha_\text{asy} = (A-Z)/A$ in Table~\ref{table:skins_normalized}, it may seem reasonable that they have similar neutron
skins. However, consider Fig.~\ref{fig:dist_comp}, which is a comparison of the neutron and proton distributions in $^{48}$Ca and $^{208}$Pb. Even normalized by particle number, the particle distributions in
$^{208}$Pb and $^{48}$Ca are quite distinct due to the size difference of the nuclei. In light of this, the neutron skin of $^{208}$Pb is biased to be larger by the increase in the rms radii of the proton and
neutron distributions. Thus, a more interesting comparison can be made by normalizing
$\Delta r_{np}$ by $r_p$, 
\begin{equation}
   \Delta\tilde{r}_{np} = \frac{1}{r_p}\Delta r_{np} = \frac{r_n}{r_p}-1,
   \label{eq:skin_normalized}
\end{equation}
where $\Delta\tilde{r}_{np}$ is the normalized neutron skin thickness.  This normalization serves to remove size dependence when comparing neutron skins of different nuclei.  The result of this normalization
is shown in Table~\ref{table:skins_normalized}. The difference between the normalized skins of $^{208}$Pb and $^{48}$Ca in Table~\ref{table:skins_normalized} reveals that the rms radius of the neutron
distribution does not simply scale by the size of the nucleus for nuclei with similar asymmetries. While it is true that the nuclear charge radius scales roughly by $A^{1/3}$ (and by extension so does $r_p$), the
same cannot be said about $r_n$. 

If one is to scale by the size of the nucleus, then the extension of the proton distribution due to Coulomb repulsion (which scales with the number of protons) should also be considered. Since $^{208}$Pb has
four times as many protons as $^{48}$Ca, the effect of Coulomb repulsion on the neutron skin of $^{208}$Pb could be up to four times more than its effect on the $^{48}$Ca neutron skin, which can reasonably
be taken from the predicted skin of $-0.06$~fm in $^{40}$Ca. In order to further investigate the effects of the Coulomb force on the neutron skin, we removed the Coulomb potential from the DOM self-energy.  In
doing this, the quasihole energy levels become much more bound, which increases the number of protons. To account for this, we shifted $\varepsilon_F$ such that it remains between the particle-hole gap of the
protons in $^{208}$Pb, corresponding to a shift of 19~MeV. Removing the effects of the Coulomb potential leads to an increased neutron skin of 0.38~fm. The results of the normalized neutron skins with Coulomb
removed are listed in Table~\ref{table:skins_normalized} for each nucleus, where it is clear that the Coulomb potential has a strong effect on the neutron skin. This points to the fact that the formation of a
neutron skin cannot be explained by the asymmetry alone. Whereas the asymmetry in $^{48}$Ca is primarily caused by the additional neutrons in the f$\frac{7}{2}$ shell, there are several different
additional shell fillings between the neutrons and protons in $^{208}$Pb. 
It is evident that these shell effects make it more difficult to predict the formation of the neutron skin based on macroscopic properties alone.

\begin{table}[tb]
   \caption[DOM Predicted neutron skins for $^{40}$Ca, $^{48}$Ca, and
   $^{208}$Pb.]
   {DOM Predicted neutron skins for $^{40}$Ca, $^{48}$Ca, and
   $^{208}$Pb. Also shown are the neutron skins normalized by $r_p$, denoted as $\Delta\tilde{r}_{np}$, as well as neutron skins with the Coulomb potential removed from the self-energy, denoted as $\Delta r^{noC}_{np}$. The last entry is the normalized neutron skin with Coulomb removed, $\Delta\tilde{r}^{noC}_{np}$.}
   \label{table:skins_normalized}
   \begin{center}
      {\renewcommand{\arraystretch}{1.15}
      \begin{tabular}{cccc}
         \hline
         \hline
         Nucleus  &  $^{40}$Ca & $^{48}$Ca & $^{208}$Pb \\
         \hline
         $\alpha_\text{asy}$ & 0 & 0.167   & 0.211  \\
         \hline
         $r_p$ & $3.39$~fm &  $3.38$~fm  &  $5.45$~fm \\
         \hline
         $r_n$ & $3.33$~fm  &  $3.63 \pm 0.023$~fm  &  $5.70 \pm 0.05$~fm \\
         \hline
         $\Delta r_{np}$   & $-0.06$~fm  & $0.25\pm0.023$~fm & $0.25\pm0.05$~fm \\
         \hline
         $\Delta\tilde{r}_{np}$ & $-0.017$ & $0.070\pm0.0067$ & $0.046\pm0.0092$ \\
         \hline
         $\Delta r^{noC}_{np}$   & $0$~fm  & $0.309\pm0.023$~fm & $0.380\pm0.05$~fm \\
         \hline
         $\Delta\tilde{r}^{noC}_{np}$ & $0$ & $0.089\pm0.0067$ & $0.070\pm0.0092$ \\
         \hline
         \hline
      \end{tabular}
   }
\end{center}
\end{table}

\begin{figure}[tb]
   \begin{minipage}{\linewidth}
      \makebox[\linewidth]{
         \includegraphics[scale=1.0]{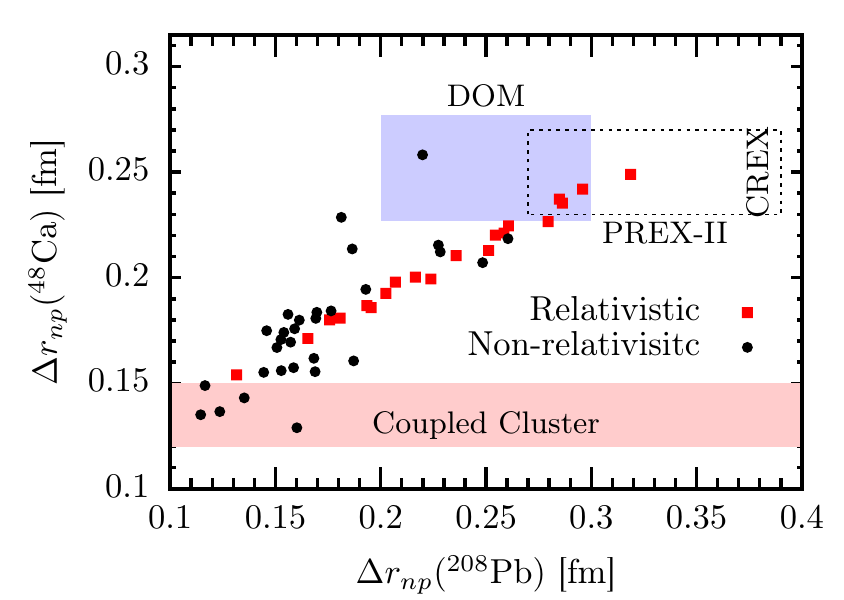}
      }
   \end{minipage}
   \caption{
   Figure adapted from Ref.~\cite{Horowitz14} with the result from Ref.~\cite{Hagen:2016} for $^{48}$Ca represented by a horizontal bar.  The shaded rectangle includes the DOM results for ${}^{208}$Pb and $^{48}$Ca~\cite{Mahzoon:2017}. Smaller squares and circles refer to relativistic and nonrelativistic mean-field calculations cited in Ref.~\cite{Horowitz14}.
   The dashed rectangle is arbitrarily centered on the DOM $^{48}$Ca result and the original PREX result (0.33), but with
updated errors of PREX-II.
   }
   \label{fig:skinc}
\end{figure}
In Fig.~\ref{fig:skinc} we present both the DOM results for $^{48}$Ca~\cite{Mahzoon:2017} and the current one for $^{208}$Pb represented by a shaded rectangle. 
The figure is adapted from Ref.~\cite{Horowitz14} and includes the coupled-cluster result from Ref.~\cite{Hagen:2016} as a horizontal band.
Relativistic and nonrelativistic mean-field calculations cited in Ref.~\cite{Horowitz14} are represented by squares and circles, respectively.
 The dashed rectangle is arbitrarily centered on the DOM $^{48}$Ca result but with the expected error of the CREX experiment~\cite{CREX13} and the original PREX result (0.33), but with updated errors expected for PREX-II~\cite{PREX-II}. 

\section{Conclusions}
\label{sec:conclusions}

We have performed a nonlocal dispersive optical-model analysis of $^{208}$Pb in which we fit elastic-scattering angular distributions, absorption and total cross sections, single-particle energies, the charge density, total binding energy, and particle number. With our well-constrained self-energy we derive a non-negliglible high-momentum content, which is consistent with the experimental observations at JLAB~\cite{CLAS:2006,Duer:2018,Hen:2017}.
Spectroscopic factors are automatically generated and appear consistent with the most up to date analysis of the $(e,e'p)$ reaction for the last valence proton orbit~\cite{Ingo91}.
Furthermore, these spectroscopic factors explain the reduction of the form factors of high spin states obtained in inelastic electron scattering~\cite{Lichtenstadt79} lending support to the interpretation of Ref.~\cite{Vijay84}.

The present analysis uses a large set of data that allow a prediction of a
neutron skin of 0.25 $\pm$ 0.05 fm.
While this is consistent with the PREX experiment~\cite{PREX12}, other methods have
been used to determine the neutron skin experimentally.
These experiments have recently been critically reviewed in Ref.~\cite{Thiel:2019} (see also
Refs.~\cite{Loc:2014,Dickhoff:2019}). The main conclusion is that these other experiments involving
hadronic probes, while valuable, continue to involve implicit model dependence
that hinder the clean determination of the neutron skin.
Our current analysis furthermore provides an alternative approach to the
multitude of mean-field calculations that provide a large variety of results
for the neutron skins of ${}^{48}$Ca and ${}^{208}$Pb~\cite{Horowitz14} while also
contrasting with the \textit{ab initio} result of Ref.~\cite{Hagen:2016} for ${}^{48}$Ca.
The new experiments employing parity-violating elastic electron scattering on
these nuclei~\cite{CREX13,PREX-II} therefore remain currently the most unambiguous approach
to determine the neutron skin.
A systematic study of more nuclei with similar asymmetry, $\alpha_\text{asy}$,
to $^{208}$Pb and $^{48}$Ca would help in determining the details of the
formation of the neutron skin. This will lead to a better understanding of the
nuclear equation of state (EOS), which is
vital to proceed in the current multi-messenger era onset by the first direct
detection of a neutron star merger~\cite{LIGO:2017}.

\section{Acknowledgements}
   This work was supported by the U.S. Department of Energy, Division of
   Nuclear Physics under grant No. DE-FG02-87ER-40316 and by the U.S. National Science Foundation under grants PHY-1613362 and PHY-1912643.

\appendix
\section{Parametrization $^{208}$Pb DOM Self-Energy}
\label{Sec:param}
We provide a detailed description of the parametrization of the proton and neutron self-energies in
$^{208}$Pb used in the fits to bound and scattering data. The functional forms are equivalent to
those used for the $^{48}$Ca potential, detailed in Ref.~\cite{Atkinson:2019}. Parameters
which are allowed to be different for protons and neutrons will contain $(n,p)$ terms.  Asymmetry
terms have been added to the amplitudes of many of the components in the form
$\pm V_{(p,n)}\frac{N-Z}{A}$ where here only, the $+$ refers to protons and $-$ to neutrons.
Elsewhere, $\pm$ in superscripts and subscripts refer to above ($+$) and below ($-$) the Fermi
energy, $\varepsilon_F$. 

We use a simple Gaussian nonlocality in all instances \cite{Perey:1962} and restrict the nonlocal contributions to the HF term and to the volume and surface contributions to the imaginary part of the potential.
We write the HF self-energy term in the following form
with spin-orbit and a local  Coulomb contribution.
\begin{equation}
\Sigma_{HF}(\bm{r},\bm{r}') = \Sigma^{nl}_{HF}(\bm{r},\bm{r}') + V^{nl}_{so}(\bm{r},\bm{r}') + \delta(\bm{r}-\bm{r}') V_C(r),
\end{equation}
The nonlocal term is split into a volume and a narrower Gaussian term of opposite sign to make the final potential have a wine-bottle shape.
\begin{equation}
\Sigma_{HF}^{nl}\left( \bm{r},\bm{r}' \right) = -V_{HF}^{vol}\left( \bm{r},\bm{r}'\right) + V_{HF}^{wb}(\bm{r},\bm{r}'),
\end{equation}
where the volume term is given by
\begin{align}
   &V_{HF}^{vol}\left( \bm{r},\bm{r}' \right) =  V^{HF}_{sym} ,f \left (
   \tilde{r},r^{HF}_{(p,n)},a^{HF}_{(p,n)} \right ) \label{eq:HFv}\\
   &\times\left [ \frac{1}{1+x_{sym}} H \left( \bm{s};\beta^{vol_1}_{sym} \right) + \frac{x_{sym}}{1+x_{sym}} H \left( \bm{s};\beta^{vol_2}_{sym}\right) \right ]  \nonumber \\
   &\pm V_{(p,n)}^{HF} \frac{N-Z}{A}f \left ( \tilde{r},r^{HFaym}_{(p,n)},a^{HFasy}_{(p,n)} \right
   )\times \nonumber \\
   &\left [ \frac{1}{1+x_{sym}} H \left( \bm{s};\beta^{vol_1asy}_{(p,n)} \right)  +
   \frac{x_{sym}}{1+x_{sym}} H \left( \bm{s};\beta^{vol_2asy}_{sym}\right) \right ].\nonumber
\end{align} 
allowing for two different nonlocalities with different weights ($0 \le x_{sym} \le1$ in
Eq.~\eqref{eq:HFv}). 
With   the notation $\tilde{r} =(r+r')/2$ and $\bm{s}=\bm{r}-\bm{r}'$,
the wine-bottle ($wb$) shape is described by
\begin{equation}
V_{HF}^{wb}(\bm{r},\bm{r}') = V^{wb}_{(p,n)}  \exp{\left(- \tilde{r}^2/(\rho^{wb}_{sym})^2\right)} H \left( \bm{s};\beta^{wb}_{sym} \right ),
\label{eq:wb}
\end{equation}
where the nonlocality in Eq.~\eqref{eq:wb} is represented by a Gaussian form
\begin{equation}
H \left( \bm{s}; \beta \right) = \exp \left( - \bm{s}^2 / \beta^2 \right)/ (\pi^{3/2} \beta^3) .
\nonumber
\end{equation}
As usual, we employ Woods-Saxon form factors 
\begin{align}
f(r,r_{i},a_{i})=\left[1+\exp \left({\frac{r-r_{i}A^{1/3}}{a_{i}}%
}\right)\right]^{-1} .
\label{Eq:WS}
\end{align}
The Coulomb term is obtained from the experimental charge density distribution for $^{48}$Ca~\cite{deVries:1987}.

The local spin-orbit interaction is given by
\begin{align}
      V_{so}(\bm{r},\bm{r'})= \left( \frac{\hbar}{m_{\pi }c}\right) ^{2}
      V^{so}_{(p,n)}\frac{1}{\tilde{r}}\frac{d}{d\tilde{r}}f(\tilde{r},r^{so}_{(p,n)},a^{so})\nonumber \\
      \times\bm{\ell}\cdot \bm{\sigma}
      H(\bm{s};\beta^{so}), 
   \label{eq:HFso}
\end{align}
where $\left( \hbar /m_{\pi }c\right) ^{2}$=2.0~fm$^{2}$ 
as in Ref.~\cite{Mueller:2011}.

The fully-nonlocal imaginary part of the DOM self-energy has the following form,
\begin{align}
\label{eq:imnl}
&\textrm{Im}\ \Sigma^{nl}(\bm{r},\bm{r}';E) = \hspace{5cm} \\ \nonumber  
&-W^{vol}_{0\pm}(E) f\left(\tilde{r};r^{vol}_{\pm (p,n)};a^{vol}_{\pm}\right)H \left( \bm{s}; \beta_{\pm (p,n)}^{vol}\right) \hspace{1cm} \\ \nonumber
&+ 4a^{sur}_{sym}W^{sur0}_{\pm}\left( E\right)H \left( \bm{s}; \beta_{\pm}^{sur0}\right) \frac{d}{d \tilde{r} }f(\tilde{r},r^{sur0}_{\pm(p,n))},a^{sur}_{sym}) \\ \nonumber
& + 4 a^{sur}_{sym} W^{sur}_{\pm}(E) H \left( \bm{s};\beta^{sur}_{\pm (p,n)} \right ) \frac{d}{d \tilde{r} }f(\tilde{r},r^{sur}_{\pm(p,n)},a^{sur}_{\pm(p,n)}) \\ \nonumber 
&+ \textrm{Im}\Sigma_{so}(\bm{r},\bm{r}';E).
\end{align}
Note that the parameters relating to the shape of the imaginary spin-orbit term
are the same as those used for the real spin-orbit term.
At energies well removed
from $\varepsilon_F$, the form of the imaginary volume potential should not be
symmetric about $\varepsilon_F$ as indicated by the $\pm$ notation in the subscripts and superscripts~\cite{Dussan:2011}.
While more symmetric about $\varepsilon_F$, we have allowed a similar option for the surface absorption that is also supported by theoretical work reported in Ref.~\cite{Waldecker:2011}.
Allowing for the aforementioned asymmetry around $\varepsilon_F$ the following form was assumed for 
the depth of the volume potential~\cite{Mueller:2011}
\begin{widetext} 
\begin{equation}
W^{vol}_{0\pm}(E) =  \Delta W^{\pm}_{NM}(E) +  
\begin{cases}
0 & \text{if } |E-\varepsilon_F| < \mathcal{E}^{vol}_{\pm} \\
A^{vol}_{\pm(p,n)}  \frac{\left(|E-\varepsilon_F|-\mathcal{E}^{vol}_{\pm}\right)^4}
{\left(|E-\varepsilon_F|-\mathcal{E}^{vol}_{\pm}\right)^4 + (B^{vol}_{\pm})^4} & 
 \text{if } |E-\varepsilon_F| > \mathcal{E}^{vol}_{\pm} ,
\end{cases} 
\label{eq:volumeS}
\end{equation}
\end{widetext}
where $\Delta W^{\pm}_{NM}(E)$ in Eq.~\eqref{eq:volumeS} is the energy-asymmetric correction modeled after
nuclear-matter calculations. The asymmetry above and below $\varepsilon_F$ is essential to accommodate the Jefferson Lab $(e,e'p)$ data at large missing energy.
The energy-asymmetric correction was taken as 
\begin{widetext} 
\begin{equation}
\Delta W^{\pm}_{NM}(E)=
\begin{cases}
\alpha_{sym} A^{vol}_{+(p,n)}\left[ \sqrt{E}+\frac{\left( \varepsilon_F+\mathbb{E}_{+}\right) ^{3/2}}{2E}-\frac{3}{2}
\sqrt{\varepsilon_F+\mathbb{E}_{+}}\right] & \text{for }E-\varepsilon_F>\mathbb{E}_{+} \\ 
-  A^{vol}_{-(p,n)} \frac{(\varepsilon_F-E-\mathbb{E}_{-})^2}{(\varepsilon_F-E-\mathbb{E}_{-})^2+(\mathbb{E}_{-})^2} & \text{for }E-\varepsilon_{F}<-\mathbb{E}_{-} \\ 
0 & \text{otherwise},
\end{cases} 
\label{eq:Wnmnl}
\end{equation} 
\end{widetext}
where $E$ in Eq.~\eqref{eq:Wnmnl} corresponds to the center-of-mass energy.

To describe the energy dependence of surface absorption we employed the form of
Ref.~\cite{Charity:2007}, but include two components, one with symmetric parameters, the other with
asymmetric parameters.
\begin{align}
W^{sur0}_{\pm}\left( E\right) =\omega _{4}(E,A^{sur0}_{\pm},B^{sur0_1}_{\pm},0)- \nonumber \\
\omega_{2}(E,A^{sur0}_{\pm},B_{\pm}^{sur0_2},C^{sur0}_{\pm}),  \label{eq:paranl} 
\end{align}
\begin{align}
  W^{sur}_{\pm(p,n))}\left( E\right) =\omega _{4}(E,A^{sur}_{\pm(p,n)},B^{sur_1}_{\pm(p,n)},0)- \nonumber \\
  \omega_{2}(E,A^{sur}_{\pm(p,n)},B_{\pm(p,n)}^{sur_2},C^{sur}_{\pm(p,n)}),  \label{eq:paran2} 
\end{align}
where the $\omega$ functions in Eqs.~\eqref{eq:paranl} and~\eqref{eq:paran2} are defined as
\begin{align}
\omega _{n}(E,A^{sur},B^{sur},C^{sur})=A^{sur}\;\Theta \left(
X\right) \frac{X^{n}}{X^{n}+\left( B^{sur}\right) ^{n}}, \nonumber \\
\label{eq:omega}
\end{align}%
and $\Theta \left( X\right) $ is Heaviside's step function and $%
X=\left\vert E-\varepsilon_F\right\vert -C^{sur}$. 

The imaginary spin-orbit term in Eq.~\eqref{eq:imnl} has the same form as the real spin-orbit
term in Eq.~\eqref{eq:HFso},
\begin{align}
      W_{so}(\bm{r},\bm{r'};E)= \left( \frac{\hbar}{m_{\pi }c}\right) ^{2}
      W^{so}(E)\frac{1}{\tilde{r}}\frac{d}{d\tilde{r}}f(\tilde{r},r^{so}_{(p,n)},a^{so})\nonumber \\
      \times\bm{\ell}\cdot \bm{\sigma}
      H(\bm{s};\beta^{so}), 
   \label{eq:imag_so}
\end{align}
where the radial parameters for the imaginary component are the same as those used for the real part of the spin-orbit potential. 
It is important to note that $\textrm{Im}\Sigma_{so}$ grows with increasing $\ell$, and for large $\ell$ this can lead to an inversion of the sign of the self-energy, which results in negative occupation. While the
form of Eq.~\eqref{eq:HFso} suppresses this behavior, it is still not a proper solution. One must be careful that the magnitude of $W_{so}(E)$ does not exceed that of the volume and surface
components. As the imaginary spin-orbit component is
generally needed only at high energies, the form of Ref.~\cite{Mueller:2011} is employed, 
\begin{equation}
   W^{so}(E)= A_{sym}^{so}  \frac{(E-\varepsilon_F)^4}{(E-\varepsilon_F)^4+(B_{sym}^{so})^4} .
   \label{eq:ImSO}
\end{equation}%

With Eq.~\eqref{eq:ImSO}, all ingredients of the self energy have now been identified and their functional form described.
In addition to the Hartree-Fock contribution and the absorptive potentials we also include the
dispersive real part from all imaginary contributions according to Eq.~\eqref{eq:dispersion}.

\section*{Parameters}
\label{app:params}

\begin{table}[tb]
\caption{Parameter values for the isoscalar part of the potential. The table also contains the number of the
equation that defines each individual parameter.}
\label{Tbl:fixed}%
\begin{ruledtabular}
\begin{tabular}{ccc}
Parameter &                                  Value & Eq. \\
\hline
\multicolumn{3}{c}{Hartree-Fock} \\
\hline
$V^{HF}_{sym}$ [MeV]		& 94.0 & (\ref{eq:HFv}) \\
$a^{HF}_{sym}$ [fm]                       &  0.730   & (\ref{eq:HFv}) \\
$\beta^{vol_1}_{sym}$ [fm]       &  1.52     &(\ref{eq:HFv}) \\
$\beta^{vol_2}_{sym}$ [fm]       &  0.760     &(\ref{eq:HFv}) \\
$x_{sym}$ 				& 0.730	& (\ref{eq:HFv}) \\
$\beta^{wb}_{sym}$ [fm]                 &  0.640   & (\ref{eq:wb}) \\
\hline
\multicolumn{3}{c}{Spin-orbit} \\
\hline
$a^{so}_{sym}$ [fm]                       &  0.700   & (\ref{eq:HFso}) \\
$\beta^{so}_{sym}$ [fm]                       &  0.830   & (\ref{eq:HFso}) \\
$A^{so}_{sym}$ [MeV]                     & -3.65 & (\ref{eq:ImSO}) \\
$B^{so}_{sym}$ [MeV]                      &  208 &  (\ref{eq:ImSO})     \\
\hline
\multicolumn{3}{c}{Volume imaginary} \\
\hline
$a^{vol}_{+}$ [fm] 		& 0.470 	& (\ref{eq:imnl}) \\ 
$a^{vol}_{-}$ [fm] 		& 0.430 	& (\ref{eq:imnl}) \\ 
$\beta^{vol}_{-}$	[fm]		& 1.05	& (\ref{eq:imnl}) \\
$B^{vol}_{+}$ [MeV]		& 14.4	& (\ref{eq:volumeS}) \\
$\mathcal{E}^{vol}_{+}$ [MeV]	& 16.4	& (\ref{eq:volumeS}) \\
$B^{vol}_{-}$ [MeV]		& 84.5	& (\ref{eq:volumeS}) \\
$\mathcal{E}^{vol}_{-}$ [MeV]		& 5.50	& (\ref{eq:volumeS}) \\
$\mathbb{E}_{+}$ [MeV]			& 21.8	& (\ref{eq:Wnmnl}) \\
$\mathbb{E}_{-}$ [MeV]			& 81.1	& (\ref{eq:Wnmnl}) \\
\hline
\multicolumn{3}{c}{Surface imaginary} \\
\hline
$a^{sur}_{+sym}$ [fm] 		& 0.430 	& (\ref{eq:imnl}) \\ 
$\beta^{sur0}_{+}$ [fm]       & 1.26  & (\ref{eq:imnl}) \\ 
$a^{sur}_{-sym}$ [fm] 		& 0.550 	& (\ref{eq:imnl}) \\ 
$\beta^{sur0}_{-}$ [fm]       & 1.50  & (\ref{eq:imnl}) \\ 
$A^{sur0}_{+}$ [MeV]          & 44.2 & (\ref{eq:paranl}) \\
$B^{sur0_1}_{+}$ [MeV]         & 17.4 & (\ref{eq:paranl}) \\
$B^{sur0_2}_{+}$ [MeV]         & 24.8 & (\ref{eq:paranl}) \\
$C^{sur0}_{+}$ [MeV] & 14.0   &  (\ref{eq:paranl}) \\
$A^{sur0}_{-}$ [MeV]          & 12.6  & (\ref{eq:paranl}) \\
$B^{sur0_1}_{-}$ [MeV]         & 15.0 & (\ref{eq:paranl}) \\
$B^{sur0_2}_{-}$ [MeV]         & 80.2 & (\ref{eq:paranl}) \\
$C^{sur0}_{-}$ [MeV] & 0.950 &  (\ref{eq:paranl}) \\
\end{tabular}
\end{ruledtabular}
\end{table}
The parameters used for the symmetric part of the self-energy are presented in
Table~\ref{Tbl:fixed}. All asymmetric parameters are presented in
Table~\ref{Tbl:fitted}. 
There are 30 Lagrange-Legendre and Lagrange-Laguerre grid points used in the
$^{208}$Pb calculations~\cite{Baye_review,Baye:2010}. For $^{208}$Pb, the scaling parameter for the
Lagrange-Laguerre mesh points is $a_L=0.15$. The matching radius used for
scattering calculations is $a=12$~fm.

\begin{table}[tb]
\caption{Fitted parameter values for proton and neutron potentials in 
$^{208}$Pb. This table also lists the number of the equation that defines each individual parameter.}
\label{Tbl:fitted}%
\begin{ruledtabular}
\begin{tabular}{cccc}
Parameter &    $(p)$ Value    &  $(n)$ Value & Eq. \\
\hline
\multicolumn{4}{c}{Hartree-Fock} \\
\hline
$V^{HF}_{(p,n)}$ [MeV]  & 22.7 &  71.1  & (\ref{eq:HFv}) \\
$r^{HF}_{(p,n)}$ [fm] &  1.18 & 1.20  & (\ref{eq:HFv}) \\
$r^{HFasy}_{(p,n)}$ [fm] & 1.40 & 1.20 & (\ref{eq:HFv}) \\
$a^{HFasy}_{(p,n)}$ [fm] & 0.390 & 0.800 & (\ref{eq:HFv}) \\
$\beta^{vol_1asy}_{(p,n)}$ [fm] & 0.180 & 1.86 &(\ref{eq:HFv}) \\
$\beta^{vol_2asy}_{(p,n)}$ [fm] & 1.52 & 1.52 &(\ref{eq:HFv}) \\
$V_{(p,n)}^{wb}$ [MeV]    & 7.15            &  2.11   & (\ref{eq:wb}) \\
$\rho_{(p,n)}^{wb}$ [MeV]    & 0.750            &  4.00   & (\ref{eq:wb}) \\

\hline
\multicolumn{4}{c}{Spin-orbit} \\
\hline
$V^{so}_{(p,n)}$ [MeV]  & 11.6 &  8.47  & (\ref{eq:HFso}) \\
$r^{so}_{(p,n)}$ [fm]      & 1.65 &  0.970  & (\ref{eq:HFso}) \\ 
\hline

\multicolumn{4}{c}{Volume imaginary} \\
\hline

$A^{vol}_{+(p,n)}$ [MeV] & 6.93		& 3.01	& (\ref{eq:volumeS})\\
$A^{vol}_{-(p,n)}$ [MeV] & 57.0		& 60.4	& (\ref{eq:volumeS})\\
$B^{vol}_{+(p,n)}$ [MeV] & 14.4		& 14.4	& (\ref{eq:volumeS})\\
$B^{vol}_{-(p,n)}$ [MeV] & 84.5		& 84.5	& (\ref{eq:volumeS})\\
$\beta^{vol}_{+(p,n)}$ [fm] & 0.320  &  0.275  &(\ref{eq:imnl}) \\
$r^{vol}_{+(p,n)}$ [fm] & 1.35  &  1.26  &(\ref{eq:imnl}) \\
$r^{vol}_{-(p,n)}$ [fm] &  1.35 & 1.00 & (\ref{eq:imnl}) \\
$\alpha_{(p,n)}$ [fm] &  0.0800 & 0.360 & (\ref{eq:imnl}) \\

\hline
\multicolumn{4}{c}{Surface imaginary} \\
\hline
$\beta^{sur}_{-(p,n)}$ [fm] &  0.210 & 2.22 & (\ref{eq:imnl}) \\
$\beta^{sur}_{+(p,n)}$ [fm] &  1.44 & 2.03 & (\ref{eq:imnl}) \\
$A^{sur}_{+(p,n)}$ [MeV] & 50.0 & -6.49 & (\ref{eq:paran2}) \\
$A^{sur}_{-(p,n)}$ [MeV] & 0.760 & -13.0 & (\ref{eq:paran2}) \\
$B^{sur_1}_{+(p,n)}$ [MeV] & 27.7 & 18.1 & (\ref{eq:paran2}) \\
$B^{sur_2}_{+(p,n)}$ [MeV] & 60.5 & 2.40 & (\ref{eq:paran2}) \\
$C^{sur}_{+(p,n)}$ [MeV] & 200 &25.1 & (\ref{eq:paran2}) \\
$B^{sur_1}_{-(p,n)}$ [MeV] & 6.18    & 20.2 &    (\ref{eq:paran2}) \\
$B^{sur_2}_{-(p,n)}$ [MeV] & 34.3    & 40.0 &  (\ref{eq:paran2}) \\
$C^{sur}_{-(p,n)}$ [MeV] & 22.9 &1.00 & (\ref{eq:paran2}) \\
$r^{sur0}_{-(p,n)}$ [fm] & 0.970 & 0.950 &(\ref{eq:imnl}) \\
$r^{sur0}_{+(p,n)}$ [fm] & 1.09 & 1.35 &(\ref{eq:imnl}) \\
$r^{sur}_{-(p,n)}$ [fm] & 0.860 & 0.860 &(\ref{eq:imnl}) \\
$r^{sur}_{+(p,n)}$ [fm] & 1.20 & 1.630 &(\ref{eq:imnl}) \\
$a^{sur}_{-(p,n)}$ [fm] & 0.600     & 0.600 &  (\ref{eq:imnl}) \\
$a^{sur}_{+(p,n)}$ [fm] & 0.530     & 0.470 &  (\ref{eq:imnl}) \\

\end{tabular}
\end{ruledtabular}
\end{table}

\newpage

\bibliographystyle{apsrev4-1}
\bibliography{pb208}

\begin{thebibliography}{72}%
\makeatletter
\providecommand \@ifxundefined [1]{%
 \@ifx{#1\undefined}
}%
\providecommand \@ifnum [1]{%
 \ifnum #1\expandafter \@firstoftwo
 \else \expandafter \@secondoftwo
 \fi
}%
\providecommand \@ifx [1]{%
 \ifx #1\expandafter \@firstoftwo
 \else \expandafter \@secondoftwo
 \fi
}%
\providecommand \natexlab [1]{#1}%
\providecommand \enquote  [1]{``#1''}%
\providecommand \bibnamefont  [1]{#1}%
\providecommand \bibfnamefont [1]{#1}%
\providecommand \citenamefont [1]{#1}%
\providecommand \href@noop [0]{\@secondoftwo}%
\providecommand \href [0]{\begingroup \@sanitize@url \@href}%
\providecommand \@href[1]{\@@startlink{#1}\@@href}%
\providecommand \@@href[1]{\endgroup#1\@@endlink}%
\providecommand \@sanitize@url [0]{\catcode `\\12\catcode `\$12\catcode
  `\&12\catcode `\#12\catcode `\^12\catcode `\_12\catcode `\%12\relax}%
\providecommand \@@startlink[1]{}%
\providecommand \@@endlink[0]{}%
\providecommand \url  [0]{\begingroup\@sanitize@url \@url }%
\providecommand \@url [1]{\endgroup\@href {#1}{\urlprefix }}%
\providecommand \urlprefix  [0]{URL }%
\providecommand \Eprint [0]{\href }%
\providecommand \doibase [0]{http://dx.doi.org/}%
\providecommand \selectlanguage [0]{\@gobble}%
\providecommand \bibinfo  [0]{\@secondoftwo}%
\providecommand \bibfield  [0]{\@secondoftwo}%
\providecommand \translation [1]{[#1]}%
\providecommand \BibitemOpen [0]{}%
\providecommand \bibitemStop [0]{}%
\providecommand \bibitemNoStop [0]{.\EOS\space}%
\providecommand \EOS [0]{\spacefactor3000\relax}%
\providecommand \BibitemShut  [1]{\csname bibitem#1\endcsname}%
\let\auto@bib@innerbib\@empty
\bibitem [{\citenamefont {Frois}\ \emph {et~al.}(1977)\citenamefont {Frois},
  \citenamefont {Bellicard}, \citenamefont {Cavedon}, \citenamefont {Huet},
  \citenamefont {Leconte}, \citenamefont {Ludeau}, \citenamefont {Nakada},
  \citenamefont {H\^o},\ and\ \citenamefont {Sick}}]{Frois77}%
  \BibitemOpen
  \bibfield  {author} {\bibinfo {author} {\bibfnamefont {B.}~\bibnamefont
  {Frois}}, \bibinfo {author} {\bibfnamefont {J.~B.}\ \bibnamefont
  {Bellicard}}, \bibinfo {author} {\bibfnamefont {J.~M.}\ \bibnamefont
  {Cavedon}}, \bibinfo {author} {\bibfnamefont {M.}~\bibnamefont {Huet}},
  \bibinfo {author} {\bibfnamefont {P.}~\bibnamefont {Leconte}}, \bibinfo
  {author} {\bibfnamefont {P.}~\bibnamefont {Ludeau}}, \bibinfo {author}
  {\bibfnamefont {A.}~\bibnamefont {Nakada}}, \bibinfo {author} {\bibfnamefont
  {P.~Z.}\ \bibnamefont {H\^o}}, \ and\ \bibinfo {author} {\bibfnamefont
  {I.}~\bibnamefont {Sick}},\ }\href {\doibase 10.1103/PhysRevLett.38.152}
  {\bibfield  {journal} {\bibinfo  {journal} {Phys. Rev. Lett.}\ }\textbf
  {\bibinfo {volume} {38}},\ \bibinfo {pages} {152} (\bibinfo {year}
  {1977})}\BibitemShut {NoStop}%
\bibitem [{\citenamefont {{de Vries}}\ \emph {et~al.}(1987)\citenamefont {{de
  Vries}}, \citenamefont {{de Jager}},\ and\ \citenamefont {{de
  Vries}}}]{deVries:1987}%
  \BibitemOpen
  \bibfield  {author} {\bibinfo {author} {\bibfnamefont {H.}~\bibnamefont {{de
  Vries}}}, \bibinfo {author} {\bibfnamefont {C.~W.}\ \bibnamefont {{de
  Jager}}}, \ and\ \bibinfo {author} {\bibfnamefont {C.}~\bibnamefont {{de
  Vries}}},\ }\href@noop {} {\bibfield  {journal} {\bibinfo  {journal} {At.
  Data Nucl. Data Tables}\ }\textbf {\bibinfo {volume} {36}},\ \bibinfo {pages}
  {495} (\bibinfo {year} {1987})}\BibitemShut {NoStop}%
\bibitem [{\citenamefont {Quint}\ \emph {et~al.}(1986)\citenamefont {Quint},
  \citenamefont {van~den Brand}, \citenamefont {den Herder}, \citenamefont
  {Jans}, \citenamefont {Keizer}, \citenamefont {Lapik\'as}, \citenamefont
  {van~der Steenhoven}, \citenamefont {de~Witt~Huberts}, \citenamefont {Klein},
  \citenamefont {Grabmayr}, \citenamefont {Wagner}, \citenamefont {Nann},
  \citenamefont {Frois},\ and\ \citenamefont {Goutte}}]{Quint86}%
  \BibitemOpen
  \bibfield  {author} {\bibinfo {author} {\bibfnamefont {E.~N.~M.}\
  \bibnamefont {Quint}}, \bibinfo {author} {\bibfnamefont {J.~F.~J.}\
  \bibnamefont {van~den Brand}}, \bibinfo {author} {\bibfnamefont {J.~W.~A.}\
  \bibnamefont {den Herder}}, \bibinfo {author} {\bibfnamefont
  {E.}~\bibnamefont {Jans}}, \bibinfo {author} {\bibfnamefont {P.~H.~M.}\
  \bibnamefont {Keizer}}, \bibinfo {author} {\bibfnamefont {L.}~\bibnamefont
  {Lapik\'as}}, \bibinfo {author} {\bibfnamefont {G.}~\bibnamefont {van~der
  Steenhoven}}, \bibinfo {author} {\bibfnamefont {P.~K.~A.}\ \bibnamefont
  {de~Witt~Huberts}}, \bibinfo {author} {\bibfnamefont {S.}~\bibnamefont
  {Klein}}, \bibinfo {author} {\bibfnamefont {P.}~\bibnamefont {Grabmayr}},
  \bibinfo {author} {\bibfnamefont {G.~J.}\ \bibnamefont {Wagner}}, \bibinfo
  {author} {\bibfnamefont {H.}~\bibnamefont {Nann}}, \bibinfo {author}
  {\bibfnamefont {B.}~\bibnamefont {Frois}}, \ and\ \bibinfo {author}
  {\bibfnamefont {D.}~\bibnamefont {Goutte}},\ }\href {\doibase
  10.1103/PhysRevLett.57.186} {\bibfield  {journal} {\bibinfo  {journal} {Phys.
  Rev. Lett.}\ }\textbf {\bibinfo {volume} {57}},\ \bibinfo {pages} {186}
  (\bibinfo {year} {1986})}\BibitemShut {NoStop}%
\bibitem [{\citenamefont {Quint}\ \emph {et~al.}(1987)\citenamefont {Quint},
  \citenamefont {Barnett}, \citenamefont {van~den Berg}, \citenamefont {van~den
  Brand}, \citenamefont {Clement}, \citenamefont {Ent}, \citenamefont {Frois},
  \citenamefont {Goutte}, \citenamefont {Grabmayr}, \citenamefont {den Herder},
  \citenamefont {Jans}, \citenamefont {Kramer}, \citenamefont {Lanen},
  \citenamefont {Lapik\'as}, \citenamefont {Nann}, \citenamefont {van~der
  Steenhoven}, \citenamefont {Wagner},\ and\ \citenamefont
  {de~Witt~Huberts}}]{Quint87}%
  \BibitemOpen
  \bibfield  {author} {\bibinfo {author} {\bibfnamefont {E.~N.~M.}\
  \bibnamefont {Quint}}, \bibinfo {author} {\bibfnamefont {B.~M.}\ \bibnamefont
  {Barnett}}, \bibinfo {author} {\bibfnamefont {A.~M.}\ \bibnamefont {van~den
  Berg}}, \bibinfo {author} {\bibfnamefont {J.~F.~J.}\ \bibnamefont {van~den
  Brand}}, \bibinfo {author} {\bibfnamefont {H.}~\bibnamefont {Clement}},
  \bibinfo {author} {\bibfnamefont {R.}~\bibnamefont {Ent}}, \bibinfo {author}
  {\bibfnamefont {B.}~\bibnamefont {Frois}}, \bibinfo {author} {\bibfnamefont
  {D.}~\bibnamefont {Goutte}}, \bibinfo {author} {\bibfnamefont
  {P.}~\bibnamefont {Grabmayr}}, \bibinfo {author} {\bibfnamefont {J.~W.~A.}\
  \bibnamefont {den Herder}}, \bibinfo {author} {\bibfnamefont
  {E.}~\bibnamefont {Jans}}, \bibinfo {author} {\bibfnamefont {G.~J.}\
  \bibnamefont {Kramer}}, \bibinfo {author} {\bibfnamefont {J.~B. J.~M.}\
  \bibnamefont {Lanen}}, \bibinfo {author} {\bibfnamefont {L.}~\bibnamefont
  {Lapik\'as}}, \bibinfo {author} {\bibfnamefont {H.}~\bibnamefont {Nann}},
  \bibinfo {author} {\bibfnamefont {G.}~\bibnamefont {van~der Steenhoven}},
  \bibinfo {author} {\bibfnamefont {G.~J.}\ \bibnamefont {Wagner}}, \ and\
  \bibinfo {author} {\bibfnamefont {P.~K.~A.}\ \bibnamefont
  {de~Witt~Huberts}},\ }\href {\doibase 10.1103/PhysRevLett.58.1088} {\bibfield
   {journal} {\bibinfo  {journal} {Phys. Rev. Lett.}\ }\textbf {\bibinfo
  {volume} {58}},\ \bibinfo {pages} {1088} (\bibinfo {year}
  {1987})}\BibitemShut {NoStop}%
\bibitem [{\citenamefont {Quint}(1988)}]{Quint88}%
  \BibitemOpen
  \bibfield  {author} {\bibinfo {author} {\bibfnamefont {E.~N.~M.}\
  \bibnamefont {Quint}},\ }\href@noop {} {Ph.D. thesis},\ \bibinfo  {school}
  {Universiteit van Amsterdam, Amsterdam} (\bibinfo {year} {1988})\BibitemShut
  {NoStop}%
\bibitem [{\citenamefont {Lichtenstadt}\ \emph {et~al.}(1979)\citenamefont
  {Lichtenstadt}, \citenamefont {Heisenberg}, \citenamefont {Papanicolas},
  \citenamefont {Sargent}, \citenamefont {Courtemanche},\ and\ \citenamefont
  {McCarthy}}]{Lichtenstadt79}%
  \BibitemOpen
  \bibfield  {author} {\bibinfo {author} {\bibfnamefont {J.}~\bibnamefont
  {Lichtenstadt}}, \bibinfo {author} {\bibfnamefont {J.}~\bibnamefont
  {Heisenberg}}, \bibinfo {author} {\bibfnamefont {C.~N.}\ \bibnamefont
  {Papanicolas}}, \bibinfo {author} {\bibfnamefont {C.~P.}\ \bibnamefont
  {Sargent}}, \bibinfo {author} {\bibfnamefont {A.~N.}\ \bibnamefont
  {Courtemanche}}, \ and\ \bibinfo {author} {\bibfnamefont {J.~S.}\
  \bibnamefont {McCarthy}},\ }\href {\doibase 10.1103/PhysRevC.20.497}
  {\bibfield  {journal} {\bibinfo  {journal} {Phys. Rev. C}\ }\textbf {\bibinfo
  {volume} {20}},\ \bibinfo {pages} {497} (\bibinfo {year} {1979})}\BibitemShut
  {NoStop}%
\bibitem [{\citenamefont {Pandharipande}\ \emph {et~al.}(1984)\citenamefont
  {Pandharipande}, \citenamefont {Papanicolas},\ and\ \citenamefont
  {Wambach}}]{Vijay84}%
  \BibitemOpen
  \bibfield  {author} {\bibinfo {author} {\bibfnamefont {V.~R.}\ \bibnamefont
  {Pandharipande}}, \bibinfo {author} {\bibfnamefont {C.~N.}\ \bibnamefont
  {Papanicolas}}, \ and\ \bibinfo {author} {\bibfnamefont {J.}~\bibnamefont
  {Wambach}},\ }\href {\doibase 10.1103/PhysRevLett.53.1133} {\bibfield
  {journal} {\bibinfo  {journal} {Phys. Rev. Lett.}\ }\textbf {\bibinfo
  {volume} {53}},\ \bibinfo {pages} {1133} (\bibinfo {year}
  {1984})}\BibitemShut {NoStop}%
\bibitem [{\citenamefont {Hen}\ \emph {et~al.}(2017)\citenamefont {Hen},
  \citenamefont {Miller}, \citenamefont {Piasetzky},\ and\ \citenamefont
  {Weinstein}}]{Hen:2017}%
  \BibitemOpen
  \bibfield  {author} {\bibinfo {author} {\bibfnamefont {O.}~\bibnamefont
  {Hen}}, \bibinfo {author} {\bibfnamefont {G.~A.}\ \bibnamefont {Miller}},
  \bibinfo {author} {\bibfnamefont {E.}~\bibnamefont {Piasetzky}}, \ and\
  \bibinfo {author} {\bibfnamefont {L.~B.}\ \bibnamefont {Weinstein}},\ }\href
  {\doibase 10.1103/RevModPhys.89.045002} {\bibfield  {journal} {\bibinfo
  {journal} {Rev. Mod. Phys.}\ }\textbf {\bibinfo {volume} {89}},\ \bibinfo
  {pages} {045002} (\bibinfo {year} {2017})}\BibitemShut {NoStop}%
\bibitem [{\citenamefont {Duer}\ \emph {et~al.}(2018)\citenamefont {Duer} \emph
  {et~al.}}]{Duer:2018}%
  \BibitemOpen
  \bibfield  {author} {\bibinfo {author} {\bibfnamefont {M.}~\bibnamefont
  {Duer}} \emph {et~al.},\ }\href {\doibase 10.1038/s41586-018-0400-z}
  {\bibfield  {journal} {\bibinfo  {journal} {Nature}\ }\textbf {\bibinfo
  {volume} {560}},\ \bibinfo {pages} {617} (\bibinfo {year}
  {2018})}\BibitemShut {NoStop}%
\bibitem [{\citenamefont {Dickhoff}\ and\ \citenamefont
  {Barbieri}(2004)}]{Dickhoff04}%
  \BibitemOpen
  \bibfield  {author} {\bibinfo {author} {\bibfnamefont {W.~H.}\ \bibnamefont
  {Dickhoff}}\ and\ \bibinfo {author} {\bibfnamefont {C.}~\bibnamefont
  {Barbieri}},\ }\href {\doibase 10.1016/j.ppnp.2004.02.038} {\bibfield
  {journal} {\bibinfo  {journal} {Prog. Part. Nucl. Phys.}\ }\textbf {\bibinfo
  {volume} {52}},\ \bibinfo {pages} {377} (\bibinfo {year} {2004})}\BibitemShut
  {NoStop}%
\bibitem [{\citenamefont {Mahaux}\ and\ \citenamefont
  {Sartor}(1991)}]{Mahaux:91}%
  \BibitemOpen
  \bibfield  {author} {\bibinfo {author} {\bibfnamefont {C.}~\bibnamefont
  {Mahaux}}\ and\ \bibinfo {author} {\bibfnamefont {R.}~\bibnamefont
  {Sartor}},\ }in\ \href {\doibase 10.1007/978-1-4613-9910-0_1} {\emph
  {\bibinfo {booktitle} {Adv. Nucl. Phys.}}},\ Vol.~\bibinfo {volume} {20}\
  (\bibinfo  {publisher} {Springer US},\ \bibinfo {year} {1991})\ p.~\bibinfo
  {pages} {1}\BibitemShut {NoStop}%
\bibitem [{\citenamefont {Dickhoff}\ \emph {et~al.}(2017)\citenamefont
  {Dickhoff}, \citenamefont {Charity},\ and\ \citenamefont
  {Mahzoon}}]{Dickhoff:2017}%
  \BibitemOpen
  \bibfield  {author} {\bibinfo {author} {\bibfnamefont {W.~H.}\ \bibnamefont
  {Dickhoff}}, \bibinfo {author} {\bibfnamefont {R.~J.}\ \bibnamefont
  {Charity}}, \ and\ \bibinfo {author} {\bibfnamefont {M.~H.}\ \bibnamefont
  {Mahzoon}},\ }\href {http://stacks.iop.org/0954-3899/44/i=3/a=033001}
  {\bibfield  {journal} {\bibinfo  {journal} {J. of Phys. G: Nucl. and Part.
  Phys.}\ }\textbf {\bibinfo {volume} {44}},\ \bibinfo {pages} {033001}
  (\bibinfo {year} {2017})}\BibitemShut {NoStop}%
\bibitem [{\citenamefont {Dickhoff}\ and\ \citenamefont
  {Charity}(2019)}]{Dickhoff:2019}%
  \BibitemOpen
  \bibfield  {author} {\bibinfo {author} {\bibfnamefont {W.~H.}\ \bibnamefont
  {Dickhoff}}\ and\ \bibinfo {author} {\bibfnamefont {R.~J.}\ \bibnamefont
  {Charity}},\ }\href {\doibase https://doi.org/10.1016/j.ppnp.2018.11.002}
  {\bibfield  {journal} {\bibinfo  {journal} {Prog. Part. Nucl. Phys.}\
  }\textbf {\bibinfo {volume} {105}},\ \bibinfo {pages} {252 } (\bibinfo {year}
  {2019})}\BibitemShut {NoStop}%
\bibitem [{\citenamefont {Dickhoff}\ and\ \citenamefont {{Van
  Neck}}(2008)}]{Exposed!}%
  \BibitemOpen
  \bibfield  {author} {\bibinfo {author} {\bibfnamefont {W.~H.}\ \bibnamefont
  {Dickhoff}}\ and\ \bibinfo {author} {\bibfnamefont {D.}~\bibnamefont {{Van
  Neck}}},\ }\href@noop {} {\emph {\bibinfo {title} {Many-Body Theory Exposed!,
  2nd edition}}}\ (\bibinfo  {publisher} {World Scientific},\ \bibinfo
  {address} {New Jersey},\ \bibinfo {year} {2008})\BibitemShut {NoStop}%
\bibitem [{\citenamefont {Mahzoon}\ \emph {et~al.}(2014)\citenamefont
  {Mahzoon}, \citenamefont {Charity}, \citenamefont {Dickhoff}, \citenamefont
  {Dussan},\ and\ \citenamefont {Waldecker}}]{Mahzoon:2014}%
  \BibitemOpen
  \bibfield  {author} {\bibinfo {author} {\bibfnamefont {M.~H.}\ \bibnamefont
  {Mahzoon}}, \bibinfo {author} {\bibfnamefont {R.~J.}\ \bibnamefont
  {Charity}}, \bibinfo {author} {\bibfnamefont {W.~H.}\ \bibnamefont
  {Dickhoff}}, \bibinfo {author} {\bibfnamefont {H.}~\bibnamefont {Dussan}}, \
  and\ \bibinfo {author} {\bibfnamefont {S.~J.}\ \bibnamefont {Waldecker}},\
  }\href {\doibase 10.1103/PhysRevLett.112.162503} {\bibfield  {journal}
  {\bibinfo  {journal} {Phys. Rev. Lett.}\ }\textbf {\bibinfo {volume} {112}},\
  \bibinfo {pages} {162503} (\bibinfo {year} {2014})}\BibitemShut {NoStop}%
\bibitem [{\citenamefont {Dussan}\ \emph {et~al.}(2014)\citenamefont {Dussan},
  \citenamefont {Mahzoon}, \citenamefont {Charity}, \citenamefont {Dickhoff},\
  and\ \citenamefont {Polls}}]{Dussan:2014}%
  \BibitemOpen
  \bibfield  {author} {\bibinfo {author} {\bibfnamefont {H.}~\bibnamefont
  {Dussan}}, \bibinfo {author} {\bibfnamefont {M.~H.}\ \bibnamefont {Mahzoon}},
  \bibinfo {author} {\bibfnamefont {R.~J.}\ \bibnamefont {Charity}}, \bibinfo
  {author} {\bibfnamefont {W.~H.}\ \bibnamefont {Dickhoff}}, \ and\ \bibinfo
  {author} {\bibfnamefont {A.}~\bibnamefont {Polls}},\ }\href {\doibase
  10.1103/PhysRevC.90.061603} {\bibfield  {journal} {\bibinfo  {journal} {Phys.
  Rev. C}\ }\textbf {\bibinfo {volume} {90}},\ \bibinfo {pages} {061603}
  (\bibinfo {year} {2014})}\BibitemShut {NoStop}%
\bibitem [{\citenamefont {Atkinson}\ \emph {et~al.}(2018)\citenamefont
  {Atkinson}, \citenamefont {Blok}, \citenamefont {Lapik\'as}, \citenamefont
  {Charity},\ and\ \citenamefont {Dickhoff}}]{Atkinson:2018}%
  \BibitemOpen
  \bibfield  {author} {\bibinfo {author} {\bibfnamefont {M.~C.}\ \bibnamefont
  {Atkinson}}, \bibinfo {author} {\bibfnamefont {H.~P.}\ \bibnamefont {Blok}},
  \bibinfo {author} {\bibfnamefont {L.}~\bibnamefont {Lapik\'as}}, \bibinfo
  {author} {\bibfnamefont {R.~J.}\ \bibnamefont {Charity}}, \ and\ \bibinfo
  {author} {\bibfnamefont {W.~H.}\ \bibnamefont {Dickhoff}},\ }\href {\doibase
  10.1103/PhysRevC.98.044627} {\bibfield  {journal} {\bibinfo  {journal} {Phys.
  Rev. C}\ }\textbf {\bibinfo {volume} {98}},\ \bibinfo {pages} {044627}
  (\bibinfo {year} {2018})}\BibitemShut {NoStop}%
\bibitem [{\citenamefont {Lapik{\'a}s}(1993)}]{Lapikas93}%
  \BibitemOpen
  \bibfield  {author} {\bibinfo {author} {\bibfnamefont {L.}~\bibnamefont
  {Lapik{\'a}s}},\ }\href {\doibase 10.1016/0375-9474(93)90630-G} {\bibfield
  {journal} {\bibinfo  {journal} {Nucl. Phys.}\ }\textbf {\bibinfo {volume}
  {A553}},\ \bibinfo {pages} {297c} (\bibinfo {year} {1993})}\BibitemShut
  {NoStop}%
\bibitem [{\citenamefont {Atkinson}\ and\ \citenamefont
  {Dickhoff}(2019)}]{Atkinson:2019}%
  \BibitemOpen
  \bibfield  {author} {\bibinfo {author} {\bibfnamefont {M.~C.}\ \bibnamefont
  {Atkinson}}\ and\ \bibinfo {author} {\bibfnamefont {W.~H.}\ \bibnamefont
  {Dickhoff}},\ }\href {\doibase
  https://doi.org/10.1016/j.physletb.2019.135027} {\bibfield  {journal}
  {\bibinfo  {journal} {Phys. Lett. B}\ }\textbf {\bibinfo {volume} {798}},\
  \bibinfo {pages} {135027} (\bibinfo {year} {2019})}\BibitemShut {NoStop}%
\bibitem [{\citenamefont {Tostevin}\ and\ \citenamefont
  {Gade}(2014)}]{Gade:2014}%
  \BibitemOpen
  \bibfield  {author} {\bibinfo {author} {\bibfnamefont {J.~A.}\ \bibnamefont
  {Tostevin}}\ and\ \bibinfo {author} {\bibfnamefont {A.}~\bibnamefont
  {Gade}},\ }\href {\doibase 10.1103/PhysRevC.90.057602} {\bibfield  {journal}
  {\bibinfo  {journal} {Phys. Rev. C}\ }\textbf {\bibinfo {volume} {90}},\
  \bibinfo {pages} {057602} (\bibinfo {year} {2014})}\BibitemShut {NoStop}%
\bibitem [{\citenamefont {Atar}\ \emph {et~al.}(2018)\citenamefont {Atar} \emph
  {et~al.}}]{Aumann18}%
  \BibitemOpen
  \bibfield  {author} {\bibinfo {author} {\bibfnamefont {L.}~\bibnamefont
  {Atar}} \emph {et~al.},\ }\href {\doibase 10.1103/PhysRevLett.120.052501}
  {\bibfield  {journal} {\bibinfo  {journal} {Phys. Rev. Lett.}\ }\textbf
  {\bibinfo {volume} {120}},\ \bibinfo {pages} {052501} (\bibinfo {year}
  {2018})}\BibitemShut {NoStop}%
\bibitem [{\citenamefont {Kawase}\ \emph {et~al.}(2018)\citenamefont {Kawase}
  \emph {et~al.}}]{Kawase:2018}%
  \BibitemOpen
  \bibfield  {author} {\bibinfo {author} {\bibfnamefont {S.}~\bibnamefont
  {Kawase}} \emph {et~al.},\ }\href {\doibase 10.1093/ptep/pty011} {\bibfield
  {journal} {\bibinfo  {journal} {Prog. Theor. Exp. Phys.}\ }\textbf {\bibinfo
  {volume} {2018}},\ \bibinfo {pages} {021D01} (\bibinfo {year}
  {2018})}\BibitemShut {NoStop}%
\bibitem [{\citenamefont {Mahzoon}\ \emph {et~al.}(2017)\citenamefont
  {Mahzoon}, \citenamefont {Atkinson}, \citenamefont {Charity},\ and\
  \citenamefont {Dickhoff}}]{Mahzoon:2017}%
  \BibitemOpen
  \bibfield  {author} {\bibinfo {author} {\bibfnamefont {M.~H.}\ \bibnamefont
  {Mahzoon}}, \bibinfo {author} {\bibfnamefont {M.~C.}\ \bibnamefont
  {Atkinson}}, \bibinfo {author} {\bibfnamefont {R.~J.}\ \bibnamefont
  {Charity}}, \ and\ \bibinfo {author} {\bibfnamefont {W.~H.}\ \bibnamefont
  {Dickhoff}},\ }\href {\doibase 10.1103/PhysRevLett.119.222503} {\bibfield
  {journal} {\bibinfo  {journal} {Phys. Rev. Lett.}\ }\textbf {\bibinfo
  {volume} {119}},\ \bibinfo {pages} {222503} (\bibinfo {year}
  {2017})}\BibitemShut {NoStop}%
\bibitem [{\citenamefont {Abrahamyan}\ \emph {et~al.}(2012)\citenamefont
  {Abrahamyan} \emph {et~al.}}]{PREX12}%
  \BibitemOpen
  \bibfield  {author} {\bibinfo {author} {\bibfnamefont {S.}~\bibnamefont
  {Abrahamyan}} \emph {et~al.} (\bibinfo {collaboration} {PREX
  Collaboration}),\ }\href {\doibase 10.1103/PhysRevLett.108.112502} {\bibfield
   {journal} {\bibinfo  {journal} {Phys. Rev. Lett.}\ }\textbf {\bibinfo
  {volume} {108}},\ \bibinfo {pages} {112502} (\bibinfo {year}
  {2012})}\BibitemShut {NoStop}%
\bibitem [{\citenamefont {Pastore}\ \emph {et~al.}(2018)\citenamefont
  {Pastore}, \citenamefont {Carlson}, \citenamefont {Cirigliano}, \citenamefont
  {Dekens}, \citenamefont {Mereghetti},\ and\ \citenamefont
  {Wiringa}}]{Pastore:2018}%
  \BibitemOpen
  \bibfield  {author} {\bibinfo {author} {\bibfnamefont {S.}~\bibnamefont
  {Pastore}}, \bibinfo {author} {\bibfnamefont {J.}~\bibnamefont {Carlson}},
  \bibinfo {author} {\bibfnamefont {V.}~\bibnamefont {Cirigliano}}, \bibinfo
  {author} {\bibfnamefont {W.}~\bibnamefont {Dekens}}, \bibinfo {author}
  {\bibfnamefont {E.}~\bibnamefont {Mereghetti}}, \ and\ \bibinfo {author}
  {\bibfnamefont {R.~B.}\ \bibnamefont {Wiringa}},\ }\href {\doibase
  10.1103/PhysRevC.97.014606} {\bibfield  {journal} {\bibinfo  {journal} {Phys.
  Rev. C}\ }\textbf {\bibinfo {volume} {97}},\ \bibinfo {pages} {014606}
  (\bibinfo {year} {2018})}\BibitemShut {NoStop}%
\bibitem [{\citenamefont {Hyv\"arinen}\ and\ \citenamefont
  {Suhonen}(2015)}]{Hyvarinen:2015}%
  \BibitemOpen
  \bibfield  {author} {\bibinfo {author} {\bibfnamefont {J.}~\bibnamefont
  {Hyv\"arinen}}\ and\ \bibinfo {author} {\bibfnamefont {J.}~\bibnamefont
  {Suhonen}},\ }\href {\doibase 10.1103/PhysRevC.91.024613} {\bibfield
  {journal} {\bibinfo  {journal} {Phys. Rev. C}\ }\textbf {\bibinfo {volume}
  {91}},\ \bibinfo {pages} {024613} (\bibinfo {year} {2015})}\BibitemShut
  {NoStop}%
\bibitem [{\citenamefont {Typel}\ and\ \citenamefont {Brown}(2001)}]{Typel01}%
  \BibitemOpen
  \bibfield  {author} {\bibinfo {author} {\bibfnamefont {S.}~\bibnamefont
  {Typel}}\ and\ \bibinfo {author} {\bibfnamefont {B.~A.}\ \bibnamefont
  {Brown}},\ }\href {\doibase 10.1103/PhysRevC.64.027302} {\bibfield  {journal}
  {\bibinfo  {journal} {Phys. Rev. C}\ }\textbf {\bibinfo {volume} {64}},\
  \bibinfo {pages} {027302} (\bibinfo {year} {2001})}\BibitemShut {NoStop}%
\bibitem [{\citenamefont {Furnstahl}\ and\ \citenamefont
  {Hammer}(2002)}]{Furnstahl02}%
  \BibitemOpen
  \bibfield  {author} {\bibinfo {author} {\bibfnamefont {R.~J.}\ \bibnamefont
  {Furnstahl}}\ and\ \bibinfo {author} {\bibfnamefont {H.}~\bibnamefont
  {Hammer}},\ }\href@noop {} {\bibfield  {journal} {\bibinfo  {journal} {Phys.
  Lett. B}\ }\textbf {\bibinfo {volume} {531}},\ \bibinfo {pages} {203}
  (\bibinfo {year} {2002})}\BibitemShut {NoStop}%
\bibitem [{\citenamefont {Steiner}\ \emph {et~al.}(2005)\citenamefont
  {Steiner}, \citenamefont {Prakash}, \citenamefont {Lattimer},\ and\
  \citenamefont {Ellis}}]{Steiner05}%
  \BibitemOpen
  \bibfield  {author} {\bibinfo {author} {\bibfnamefont {A.}~\bibnamefont
  {Steiner}}, \bibinfo {author} {\bibfnamefont {M.}~\bibnamefont {Prakash}},
  \bibinfo {author} {\bibfnamefont {J.}~\bibnamefont {Lattimer}}, \ and\
  \bibinfo {author} {\bibfnamefont {P.}~\bibnamefont {Ellis}},\ }\href
  {\doibase http://dx.doi.org/10.1016/j.physrep.2005.02.004} {\bibfield
  {journal} {\bibinfo  {journal} {Phys. Rep.}\ }\textbf {\bibinfo {volume}
  {411}},\ \bibinfo {pages} {325 } (\bibinfo {year} {2005})}\BibitemShut
  {NoStop}%
\bibitem [{\citenamefont {Roca-Maza}\ \emph {et~al.}(2011)\citenamefont
  {Roca-Maza}, \citenamefont {Centelles}, \citenamefont {Vi\~nas},\ and\
  \citenamefont {Warda}}]{RocaMaza11}%
  \BibitemOpen
  \bibfield  {author} {\bibinfo {author} {\bibfnamefont {X.}~\bibnamefont
  {Roca-Maza}}, \bibinfo {author} {\bibfnamefont {M.}~\bibnamefont
  {Centelles}}, \bibinfo {author} {\bibfnamefont {X.}~\bibnamefont {Vi\~nas}},
  \ and\ \bibinfo {author} {\bibfnamefont {M.}~\bibnamefont {Warda}},\ }\href
  {\doibase 10.1103/PhysRevLett.106.252501} {\bibfield  {journal} {\bibinfo
  {journal} {Phys. Rev. Lett.}\ }\textbf {\bibinfo {volume} {106}},\ \bibinfo
  {pages} {252501} (\bibinfo {year} {2011})}\BibitemShut {NoStop}%
\bibitem [{\citenamefont {Horowitz}\ and\ \citenamefont
  {Piekarewicz}(2001)}]{Horowitz01}%
  \BibitemOpen
  \bibfield  {author} {\bibinfo {author} {\bibfnamefont {C.~J.}\ \bibnamefont
  {Horowitz}}\ and\ \bibinfo {author} {\bibfnamefont {J.}~\bibnamefont
  {Piekarewicz}},\ }\href {\doibase 10.1103/PhysRevLett.86.5647} {\bibfield
  {journal} {\bibinfo  {journal} {Phys. Rev. Lett.}\ }\textbf {\bibinfo
  {volume} {86}},\ \bibinfo {pages} {5647} (\bibinfo {year}
  {2001})}\BibitemShut {NoStop}%
\bibitem [{\citenamefont {Steiner}\ \emph {et~al.}(2010)\citenamefont
  {Steiner}, \citenamefont {Lattimer},\ and\ \citenamefont
  {Brown}}]{Steiner10}%
  \BibitemOpen
  \bibfield  {author} {\bibinfo {author} {\bibfnamefont {A.~W.}\ \bibnamefont
  {Steiner}}, \bibinfo {author} {\bibfnamefont {J.~M.}\ \bibnamefont
  {Lattimer}}, \ and\ \bibinfo {author} {\bibfnamefont {E.~F.}\ \bibnamefont
  {Brown}},\ }\href {http://stacks.iop.org/0004-637X/722/i=1/a=33} {\bibfield
  {journal} {\bibinfo  {journal} {Astrophys. J.}\ }\textbf {\bibinfo {volume}
  {722}},\ \bibinfo {pages} {33} (\bibinfo {year} {2010})}\BibitemShut
  {NoStop}%
\bibitem [{\citenamefont {Li}\ \emph {et~al.}(2008)\citenamefont {Li},
  \citenamefont {Chen},\ and\ \citenamefont {Ko}}]{li08}%
  \BibitemOpen
  \bibfield  {author} {\bibinfo {author} {\bibfnamefont {B.-A.}\ \bibnamefont
  {Li}}, \bibinfo {author} {\bibfnamefont {L.-W.}\ \bibnamefont {Chen}}, \ and\
  \bibinfo {author} {\bibfnamefont {C.~M.}\ \bibnamefont {Ko}},\ }\href
  {\doibase http://dx.doi.org/10.1016/j.physrep.2008.04.005} {\bibfield
  {journal} {\bibinfo  {journal} {Phys. Rep.}\ }\textbf {\bibinfo {volume}
  {464}},\ \bibinfo {pages} {113 } (\bibinfo {year} {2008})}\BibitemShut
  {NoStop}%
\bibitem [{\citenamefont {Tsang}\ \emph {et~al.}(2012)\citenamefont {Tsang},
  \citenamefont {Stone}, \citenamefont {Camera}, \citenamefont {Danielewicz},
  \citenamefont {Gandolfi}, \citenamefont {Hebeler}, \citenamefont {Horowitz},
  \citenamefont {Lee}, \citenamefont {Lynch}, \citenamefont {Kohley},
  \citenamefont {Lemmon}, \citenamefont {M\"oller}, \citenamefont {Murakami},
  \citenamefont {Riordan}, \citenamefont {Roca-Maza}, \citenamefont
  {Sammarruca}, \citenamefont {Steiner}, \citenamefont {Vida\~na},\ and\
  \citenamefont {Yennello}}]{Tsang12}%
  \BibitemOpen
  \bibfield  {author} {\bibinfo {author} {\bibfnamefont {M.~B.}\ \bibnamefont
  {Tsang}}, \bibinfo {author} {\bibfnamefont {J.~R.}\ \bibnamefont {Stone}},
  \bibinfo {author} {\bibfnamefont {F.}~\bibnamefont {Camera}}, \bibinfo
  {author} {\bibfnamefont {P.}~\bibnamefont {Danielewicz}}, \bibinfo {author}
  {\bibfnamefont {S.}~\bibnamefont {Gandolfi}}, \bibinfo {author}
  {\bibfnamefont {K.}~\bibnamefont {Hebeler}}, \bibinfo {author} {\bibfnamefont
  {C.~J.}\ \bibnamefont {Horowitz}}, \bibinfo {author} {\bibfnamefont
  {J.}~\bibnamefont {Lee}}, \bibinfo {author} {\bibfnamefont {W.~G.}\
  \bibnamefont {Lynch}}, \bibinfo {author} {\bibfnamefont {Z.}~\bibnamefont
  {Kohley}}, \bibinfo {author} {\bibfnamefont {R.}~\bibnamefont {Lemmon}},
  \bibinfo {author} {\bibfnamefont {P.}~\bibnamefont {M\"oller}}, \bibinfo
  {author} {\bibfnamefont {T.}~\bibnamefont {Murakami}}, \bibinfo {author}
  {\bibfnamefont {S.}~\bibnamefont {Riordan}}, \bibinfo {author} {\bibfnamefont
  {X.}~\bibnamefont {Roca-Maza}}, \bibinfo {author} {\bibfnamefont
  {F.}~\bibnamefont {Sammarruca}}, \bibinfo {author} {\bibfnamefont {A.~W.}\
  \bibnamefont {Steiner}}, \bibinfo {author} {\bibfnamefont {I.}~\bibnamefont
  {Vida\~na}}, \ and\ \bibinfo {author} {\bibfnamefont {S.~J.}\ \bibnamefont
  {Yennello}},\ }\href {\doibase 10.1103/PhysRevC.86.015803} {\bibfield
  {journal} {\bibinfo  {journal} {Phys. Rev. C}\ }\textbf {\bibinfo {volume}
  {86}},\ \bibinfo {pages} {015803} (\bibinfo {year} {2012})}\BibitemShut
  {NoStop}%
\bibitem [{\citenamefont {Angeli}\ and\ \citenamefont
  {Marinova}(2013)}]{Angeli:2013}%
  \BibitemOpen
  \bibfield  {author} {\bibinfo {author} {\bibfnamefont {I.}~\bibnamefont
  {Angeli}}\ and\ \bibinfo {author} {\bibfnamefont {K.}~\bibnamefont
  {Marinova}},\ }\href {\doibase http://dx.doi.org/10.1016/j.adt.2011.12.006}
  {\bibfield  {journal} {\bibinfo  {journal} {At. Data Nucl. Data Tables}\
  }\textbf {\bibinfo {volume} {99}},\ \bibinfo {pages} {69 } (\bibinfo {year}
  {2013})}\BibitemShut {NoStop}%
\bibitem [{\citenamefont {Horowitz}(1998)}]{Horowitz98}%
  \BibitemOpen
  \bibfield  {author} {\bibinfo {author} {\bibfnamefont {C.~J.}\ \bibnamefont
  {Horowitz}},\ }\href {\doibase 10.1103/PhysRevC.57.3430} {\bibfield
  {journal} {\bibinfo  {journal} {Phys. Rev. C}\ }\textbf {\bibinfo {volume}
  {57}},\ \bibinfo {pages} {3430} (\bibinfo {year} {1998})}\BibitemShut
  {NoStop}%
\bibitem [{\citenamefont {Piekarewicz}\ \emph {et~al.}(2012)\citenamefont
  {Piekarewicz}, \citenamefont {Agrawal}, \citenamefont {Col\`o}, \citenamefont
  {Nazarewicz}, \citenamefont {Paar}, \citenamefont {Reinhard}, \citenamefont
  {Roca-Maza},\ and\ \citenamefont {Vretenar}}]{Piekarewicz:2012}%
  \BibitemOpen
  \bibfield  {author} {\bibinfo {author} {\bibfnamefont {J.}~\bibnamefont
  {Piekarewicz}}, \bibinfo {author} {\bibfnamefont {B.~K.}\ \bibnamefont
  {Agrawal}}, \bibinfo {author} {\bibfnamefont {G.}~\bibnamefont {Col\`o}},
  \bibinfo {author} {\bibfnamefont {W.}~\bibnamefont {Nazarewicz}}, \bibinfo
  {author} {\bibfnamefont {N.}~\bibnamefont {Paar}}, \bibinfo {author}
  {\bibfnamefont {P.-G.}\ \bibnamefont {Reinhard}}, \bibinfo {author}
  {\bibfnamefont {X.}~\bibnamefont {Roca-Maza}}, \ and\ \bibinfo {author}
  {\bibfnamefont {D.}~\bibnamefont {Vretenar}},\ }\href {\doibase
  10.1103/PhysRevC.85.041302} {\bibfield  {journal} {\bibinfo  {journal} {Phys.
  Rev. C}\ }\textbf {\bibinfo {volume} {85}},\ \bibinfo {pages} {041302}
  (\bibinfo {year} {2012})}\BibitemShut {NoStop}%
\bibitem [{\citenamefont {Descouvemont}\ and\ \citenamefont
  {Baye}(2010)}]{Baye:2010}%
  \BibitemOpen
  \bibfield  {author} {\bibinfo {author} {\bibfnamefont {P.}~\bibnamefont
  {Descouvemont}}\ and\ \bibinfo {author} {\bibfnamefont {D.}~\bibnamefont
  {Baye}},\ }\href {http://stacks.iop.org/0034-4885/73/i=3/a=036301} {\bibfield
   {journal} {\bibinfo  {journal} {Rep. Prog. Phys.}\ }\textbf {\bibinfo
  {volume} {73}},\ \bibinfo {pages} {036301} (\bibinfo {year}
  {2010})}\BibitemShut {NoStop}%
\bibitem [{\citenamefont {Bell}\ and\ \citenamefont {Squires}(1959)}]{Bell59}%
  \BibitemOpen
  \bibfield  {author} {\bibinfo {author} {\bibfnamefont {J.~S.}\ \bibnamefont
  {Bell}}\ and\ \bibinfo {author} {\bibfnamefont {E.~J.}\ \bibnamefont
  {Squires}},\ }\href {\doibase 10.1103/PhysRevLett.3.96} {\bibfield  {journal}
  {\bibinfo  {journal} {Phys. Rev. Lett.}\ }\textbf {\bibinfo {volume} {3}},\
  \bibinfo {pages} {96} (\bibinfo {year} {1959})}\BibitemShut {NoStop}%
\bibitem [{\citenamefont {Perey}\ and\ \citenamefont
  {Buck}(1962)}]{Perey:1962}%
  \BibitemOpen
  \bibfield  {author} {\bibinfo {author} {\bibfnamefont {F.}~\bibnamefont
  {Perey}}\ and\ \bibinfo {author} {\bibfnamefont {B.}~\bibnamefont {Buck}},\
  }\href {\doibase https://doi.org/10.1016/0029-5582(62)90345-0} {\bibfield
  {journal} {\bibinfo  {journal} {Nuclear Physics}\ }\textbf {\bibinfo {volume}
  {32}},\ \bibinfo {pages} {353 } (\bibinfo {year} {1962})}\BibitemShut
  {NoStop}%
\bibitem [{\citenamefont {Charity}\ \emph {et~al.}(2006)\citenamefont
  {Charity}, \citenamefont {Sobotka},\ and\ \citenamefont
  {Dickhoff}}]{Charity06}%
  \BibitemOpen
  \bibfield  {author} {\bibinfo {author} {\bibfnamefont {R.~J.}\ \bibnamefont
  {Charity}}, \bibinfo {author} {\bibfnamefont {L.~G.}\ \bibnamefont
  {Sobotka}}, \ and\ \bibinfo {author} {\bibfnamefont {W.~H.}\ \bibnamefont
  {Dickhoff}},\ }\href {\doibase 10.1103/PhysRevLett.97.162503} {\bibfield
  {journal} {\bibinfo  {journal} {Phys. Rev. Lett.}\ }\textbf {\bibinfo
  {volume} {97}},\ \bibinfo {pages} {162503} (\bibinfo {year}
  {2006})}\BibitemShut {NoStop}%
\bibitem [{\citenamefont {Charity}\ \emph {et~al.}(2007)\citenamefont
  {Charity}, \citenamefont {Mueller}, \citenamefont {Sobotka},\ and\
  \citenamefont {Dickhoff}}]{Charity:2007}%
  \BibitemOpen
  \bibfield  {author} {\bibinfo {author} {\bibfnamefont {R.~J.}\ \bibnamefont
  {Charity}}, \bibinfo {author} {\bibfnamefont {J.~M.}\ \bibnamefont
  {Mueller}}, \bibinfo {author} {\bibfnamefont {L.~G.}\ \bibnamefont
  {Sobotka}}, \ and\ \bibinfo {author} {\bibfnamefont {W.~H.}\ \bibnamefont
  {Dickhoff}},\ }\href {\doibase 10.1103/PhysRevC.76.044314} {\bibfield
  {journal} {\bibinfo  {journal} {Phys. Rev. C}\ }\textbf {\bibinfo {volume}
  {76}},\ \bibinfo {pages} {044314} (\bibinfo {year} {2007})}\BibitemShut
  {NoStop}%
\bibitem [{\citenamefont {Mueller}\ \emph {et~al.}(2011)\citenamefont
  {Mueller}, \citenamefont {Charity}, \citenamefont {Shane}, \citenamefont
  {Sobotka}, \citenamefont {Waldecker}, \citenamefont {Dickhoff}, \citenamefont
  {Crowell}, \citenamefont {Esterline}, \citenamefont {Fallin}, \citenamefont
  {Howell}, \citenamefont {Westerfeldt}, \citenamefont {Youngs}, \citenamefont
  {Crowe},\ and\ \citenamefont {Pedroni}}]{Mueller:2011}%
  \BibitemOpen
  \bibfield  {author} {\bibinfo {author} {\bibfnamefont {J.~M.}\ \bibnamefont
  {Mueller}}, \bibinfo {author} {\bibfnamefont {R.~J.}\ \bibnamefont
  {Charity}}, \bibinfo {author} {\bibfnamefont {R.}~\bibnamefont {Shane}},
  \bibinfo {author} {\bibfnamefont {L.~G.}\ \bibnamefont {Sobotka}}, \bibinfo
  {author} {\bibfnamefont {S.~J.}\ \bibnamefont {Waldecker}}, \bibinfo {author}
  {\bibfnamefont {W.~H.}\ \bibnamefont {Dickhoff}}, \bibinfo {author}
  {\bibfnamefont {A.~S.}\ \bibnamefont {Crowell}}, \bibinfo {author}
  {\bibfnamefont {J.~H.}\ \bibnamefont {Esterline}}, \bibinfo {author}
  {\bibfnamefont {B.}~\bibnamefont {Fallin}}, \bibinfo {author} {\bibfnamefont
  {C.~R.}\ \bibnamefont {Howell}}, \bibinfo {author} {\bibfnamefont
  {C.}~\bibnamefont {Westerfeldt}}, \bibinfo {author} {\bibfnamefont
  {M.}~\bibnamefont {Youngs}}, \bibinfo {author} {\bibfnamefont {B.~J.}\
  \bibnamefont {Crowe}}, \ and\ \bibinfo {author} {\bibfnamefont {R.~S.}\
  \bibnamefont {Pedroni}},\ }\href {\doibase 10.1103/PhysRevC.83.064605}
  {\bibfield  {journal} {\bibinfo  {journal} {Phys. Rev. C}\ }\textbf {\bibinfo
  {volume} {83}},\ \bibinfo {pages} {064605} (\bibinfo {year}
  {2011})}\BibitemShut {NoStop}%
\bibitem [{\citenamefont {Dickhoff}\ \emph {et~al.}(2010)\citenamefont
  {Dickhoff}, \citenamefont {Van~Neck}, \citenamefont {Waldecker},
  \citenamefont {Charity},\ and\ \citenamefont {Sobotka}}]{Dickhoff:2010}%
  \BibitemOpen
  \bibfield  {author} {\bibinfo {author} {\bibfnamefont {W.~H.}\ \bibnamefont
  {Dickhoff}}, \bibinfo {author} {\bibfnamefont {D.}~\bibnamefont {Van~Neck}},
  \bibinfo {author} {\bibfnamefont {S.~J.}\ \bibnamefont {Waldecker}}, \bibinfo
  {author} {\bibfnamefont {R.~J.}\ \bibnamefont {Charity}}, \ and\ \bibinfo
  {author} {\bibfnamefont {L.~G.}\ \bibnamefont {Sobotka}},\ }\href {\doibase
  10.1103/PhysRevC.82.054306} {\bibfield  {journal} {\bibinfo  {journal} {Phys.
  Rev. C}\ }\textbf {\bibinfo {volume} {82}},\ \bibinfo {pages} {054306}
  (\bibinfo {year} {2010})}\BibitemShut {NoStop}%
\bibitem [{\citenamefont {Waldecker}\ \emph {et~al.}(2011)\citenamefont
  {Waldecker}, \citenamefont {Barbieri},\ and\ \citenamefont
  {Dickhoff}}]{Waldecker:2011}%
  \BibitemOpen
  \bibfield  {author} {\bibinfo {author} {\bibfnamefont {S.~J.}\ \bibnamefont
  {Waldecker}}, \bibinfo {author} {\bibfnamefont {C.}~\bibnamefont {Barbieri}},
  \ and\ \bibinfo {author} {\bibfnamefont {W.~H.}\ \bibnamefont {Dickhoff}},\
  }\href {\doibase 10.1103/PhysRevC.84.034616} {\bibfield  {journal} {\bibinfo
  {journal} {Phys. Rev. C}\ }\textbf {\bibinfo {volume} {84}},\ \bibinfo
  {pages} {034616} (\bibinfo {year} {2011})}\BibitemShut {NoStop}%
\bibitem [{\citenamefont {Dussan}\ \emph {et~al.}(2011)\citenamefont {Dussan},
  \citenamefont {Waldecker}, \citenamefont {Dickhoff}, \citenamefont
  {M\"uther},\ and\ \citenamefont {Polls}}]{Dussan:2011}%
  \BibitemOpen
  \bibfield  {author} {\bibinfo {author} {\bibfnamefont {H.}~\bibnamefont
  {Dussan}}, \bibinfo {author} {\bibfnamefont {S.~J.}\ \bibnamefont
  {Waldecker}}, \bibinfo {author} {\bibfnamefont {W.~H.}\ \bibnamefont
  {Dickhoff}}, \bibinfo {author} {\bibfnamefont {H.}~\bibnamefont {M\"uther}},
  \ and\ \bibinfo {author} {\bibfnamefont {A.}~\bibnamefont {Polls}},\ }\href
  {\doibase 10.1103/PhysRevC.84.044319} {\bibfield  {journal} {\bibinfo
  {journal} {Phys. Rev. C}\ }\textbf {\bibinfo {volume} {84}},\ \bibinfo
  {pages} {044319} (\bibinfo {year} {2011})}\BibitemShut {NoStop}%
\bibitem [{\citenamefont {Brida}\ \emph {et~al.}(2011)\citenamefont {Brida},
  \citenamefont {Pieper},\ and\ \citenamefont {Wiringa}}]{Brida11}%
  \BibitemOpen
  \bibfield  {author} {\bibinfo {author} {\bibfnamefont {I.}~\bibnamefont
  {Brida}}, \bibinfo {author} {\bibfnamefont {S.~C.}\ \bibnamefont {Pieper}}, \
  and\ \bibinfo {author} {\bibfnamefont {R.~B.}\ \bibnamefont {Wiringa}},\
  }\href@noop {} {\bibfield  {journal} {\bibinfo  {journal} {Phys. Rev. C}\
  }\textbf {\bibinfo {volume} {84}},\ \bibinfo {pages} {024319} (\bibinfo
  {year} {2011})}\BibitemShut {NoStop}%
\bibitem [{\citenamefont {Fiedeldey}(1966)}]{Fiedeldey:1966}%
  \BibitemOpen
  \bibfield  {author} {\bibinfo {author} {\bibfnamefont {H.}~\bibnamefont
  {Fiedeldey}},\ }\href {\doibase https://doi.org/10.1016/0029-5582(66)90682-1}
  {\bibfield  {journal} {\bibinfo  {journal} {Nucl. Phys.}\ }\textbf {\bibinfo
  {volume} {77}},\ \bibinfo {pages} {149 } (\bibinfo {year}
  {1966})}\BibitemShut {NoStop}%
\bibitem [{\citenamefont {Danielewicz}\ \emph {et~al.}(2017)\citenamefont
  {Danielewicz}, \citenamefont {Singh},\ and\ \citenamefont {Lee}}]{Pawel17}%
  \BibitemOpen
  \bibfield  {author} {\bibinfo {author} {\bibfnamefont {P.}~\bibnamefont
  {Danielewicz}}, \bibinfo {author} {\bibfnamefont {P.}~\bibnamefont {Singh}},
  \ and\ \bibinfo {author} {\bibfnamefont {J.}~\bibnamefont {Lee}},\
  }\href@noop {} {\bibfield  {journal} {\bibinfo  {journal} {Nucl. Phys. A}\
  }\textbf {\bibinfo {volume} {958}},\ \bibinfo {pages} {147} (\bibinfo {year}
  {2017})}\BibitemShut {NoStop}%
\bibitem [{\citenamefont {Loc}\ \emph {et~al.}(2014)\citenamefont {Loc},
  \citenamefont {Khoa},\ and\ \citenamefont {Zegers}}]{Loc:2014}%
  \BibitemOpen
  \bibfield  {author} {\bibinfo {author} {\bibfnamefont {B.~M.}\ \bibnamefont
  {Loc}}, \bibinfo {author} {\bibfnamefont {D.~T.}\ \bibnamefont {Khoa}}, \
  and\ \bibinfo {author} {\bibfnamefont {R.~G.~T.}\ \bibnamefont {Zegers}},\
  }\href {\doibase 10.1103/PhysRevC.89.024317} {\bibfield  {journal} {\bibinfo
  {journal} {Phys. Rev. C}\ }\textbf {\bibinfo {volume} {89}},\ \bibinfo
  {pages} {024317} (\bibinfo {year} {2014})}\BibitemShut {NoStop}%
\bibitem [{\citenamefont {Khoa}\ \emph {et~al.}(2007)\citenamefont {Khoa},
  \citenamefont {Than},\ and\ \citenamefont {Cuong}}]{Khoa:2007}%
  \BibitemOpen
  \bibfield  {author} {\bibinfo {author} {\bibfnamefont {D.~T.}\ \bibnamefont
  {Khoa}}, \bibinfo {author} {\bibfnamefont {H.~S.}\ \bibnamefont {Than}}, \
  and\ \bibinfo {author} {\bibfnamefont {D.~C.}\ \bibnamefont {Cuong}},\ }\href
  {\doibase 10.1103/PhysRevC.76.014603} {\bibfield  {journal} {\bibinfo
  {journal} {Phys. Rev. C}\ }\textbf {\bibinfo {volume} {76}},\ \bibinfo
  {pages} {014603} (\bibinfo {year} {2007})}\BibitemShut {NoStop}%
\bibitem [{\citenamefont {Koning}\ and\ \citenamefont
  {Delaroche}(2003)}]{Koning:2003}%
  \BibitemOpen
  \bibfield  {author} {\bibinfo {author} {\bibfnamefont {A.}~\bibnamefont
  {Koning}}\ and\ \bibinfo {author} {\bibfnamefont {J.}~\bibnamefont
  {Delaroche}},\ }\href {\doibase
  https://doi.org/10.1016/S0375-9474(02)01321-0} {\bibfield  {journal}
  {\bibinfo  {journal} {Nuclear Physics A}\ }\textbf {\bibinfo {volume}
  {713}},\ \bibinfo {pages} {231 } (\bibinfo {year} {2003})}\BibitemShut
  {NoStop}%
\bibitem [{\citenamefont {Press}\ \emph {et~al.}(1992)\citenamefont {Press},
  \citenamefont {Teukolsky}, \citenamefont {Vetterling},\ and\ \citenamefont
  {Flannery}}]{Numerical}%
  \BibitemOpen
  \bibfield  {author} {\bibinfo {author} {\bibfnamefont {W.~H.}\ \bibnamefont
  {Press}}, \bibinfo {author} {\bibfnamefont {S.~A.}\ \bibnamefont
  {Teukolsky}}, \bibinfo {author} {\bibfnamefont {W.~T.}\ \bibnamefont
  {Vetterling}}, \ and\ \bibinfo {author} {\bibfnamefont {B.~P.}\ \bibnamefont
  {Flannery}},\ }\href@noop {} {\emph {\bibinfo {title} {Numerical Recipes in
  C}}}\ (\bibinfo  {publisher} {Cambridge University Press},\ \bibinfo {year}
  {1992})\BibitemShut {NoStop}%
\bibitem [{\citenamefont {Sick}\ \emph {et~al.}(1979)\citenamefont {Sick},
  \citenamefont {Bellicard}, \citenamefont {Cavedon}, \citenamefont {Frois},
  \citenamefont {Huet}, \citenamefont {Leconte}, \citenamefont {Ho},\ and\
  \citenamefont {Platchkov}}]{Sick79}%
  \BibitemOpen
  \bibfield  {author} {\bibinfo {author} {\bibfnamefont {I.}~\bibnamefont
  {Sick}}, \bibinfo {author} {\bibfnamefont {J.~B.}\ \bibnamefont {Bellicard}},
  \bibinfo {author} {\bibfnamefont {J.~M.}\ \bibnamefont {Cavedon}}, \bibinfo
  {author} {\bibfnamefont {B.}~\bibnamefont {Frois}}, \bibinfo {author}
  {\bibfnamefont {M.}~\bibnamefont {Huet}}, \bibinfo {author} {\bibfnamefont
  {P.}~\bibnamefont {Leconte}}, \bibinfo {author} {\bibfnamefont {P.~X.}\
  \bibnamefont {Ho}}, \ and\ \bibinfo {author} {\bibfnamefont {S.}~\bibnamefont
  {Platchkov}},\ }\href@noop {} {\bibfield  {journal} {\bibinfo  {journal}
  {Phys. Lett. B}\ }\textbf {\bibinfo {volume} {88}},\ \bibinfo {pages} {245}
  (\bibinfo {year} {1979})}\BibitemShut {NoStop}%
\bibitem [{\citenamefont {Salvat}\ \emph {et~al.}(2005)\citenamefont {Salvat},
  \citenamefont {Jablonski},\ and\ \citenamefont {Powell}}]{salvat:2005}%
  \BibitemOpen
  \bibfield  {author} {\bibinfo {author} {\bibfnamefont {F.}~\bibnamefont
  {Salvat}}, \bibinfo {author} {\bibfnamefont {A.}~\bibnamefont {Jablonski}}, \
  and\ \bibinfo {author} {\bibfnamefont {C.~J.}\ \bibnamefont {Powell}},\
  }\href {\doibase https://doi.org/10.1016/j.cpc.2004.09.006} {\bibfield
  {journal} {\bibinfo  {journal} {Comput. Phys. Commun.}\ }\textbf {\bibinfo
  {volume} {165}},\ \bibinfo {pages} {157 } (\bibinfo {year}
  {2005})}\BibitemShut {NoStop}%
\bibitem [{\citenamefont {Bender}\ \emph {et~al.}(2003)\citenamefont {Bender},
  \citenamefont {Heenen},\ and\ \citenamefont {Reinhard}}]{Bender:2003}%
  \BibitemOpen
  \bibfield  {author} {\bibinfo {author} {\bibfnamefont {M.}~\bibnamefont
  {Bender}}, \bibinfo {author} {\bibfnamefont {P.-H.}\ \bibnamefont {Heenen}},
  \ and\ \bibinfo {author} {\bibfnamefont {P.-G.}\ \bibnamefont {Reinhard}},\
  }\href {\doibase 10.1103/RevModPhys.75.121} {\bibfield  {journal} {\bibinfo
  {journal} {Rev. Mod. Phys.}\ }\textbf {\bibinfo {volume} {75}},\ \bibinfo
  {pages} {121} (\bibinfo {year} {2003})}\BibitemShut {NoStop}%
\bibitem [{\citenamefont {Sick}\ and\ \citenamefont
  {de~Witt~Huberts}(1991)}]{Ingo91}%
  \BibitemOpen
  \bibfield  {author} {\bibinfo {author} {\bibfnamefont {I.}~\bibnamefont
  {Sick}}\ and\ \bibinfo {author} {\bibfnamefont {P.~K.~A.}\ \bibnamefont
  {de~Witt~Huberts}},\ }\href@noop {} {\bibfield  {journal} {\bibinfo
  {journal} {Comm. Nucl. Part. Phys.}\ }\textbf {\bibinfo {volume} {20}},\
  \bibinfo {pages} {177} (\bibinfo {year} {1991})}\BibitemShut {NoStop}%
\bibitem [{\citenamefont {{van Batenburg}}(2001)}]{Batenburg01}%
  \BibitemOpen
  \bibfield  {author} {\bibinfo {author} {\bibfnamefont {M.~F.}\ \bibnamefont
  {{van Batenburg}}},\ }\href@noop {} {\emph {\bibinfo {title} {Ph.D.
  Thesis}}}\ (\bibinfo  {publisher} {University of Utrecht},\ \bibinfo {year}
  {2001})\BibitemShut {NoStop}%
\bibitem [{\citenamefont {Egiyan}\ \emph {et~al.}(2006)\citenamefont {Egiyan}
  \emph {et~al.}}]{CLAS:2006}%
  \BibitemOpen
  \bibfield  {author} {\bibinfo {author} {\bibfnamefont {K.~S.}\ \bibnamefont
  {Egiyan}} \emph {et~al.} (\bibinfo {collaboration} {CLAS Collaboration}),\
  }\href {\doibase 10.1103/PhysRevLett.96.082501} {\bibfield  {journal}
  {\bibinfo  {journal} {Phys. Rev. Lett.}\ }\textbf {\bibinfo {volume} {96}},\
  \bibinfo {pages} {082501} (\bibinfo {year} {2006})}\BibitemShut {NoStop}%
\bibitem [{\citenamefont {Frick}\ \emph {et~al.}(2005)\citenamefont {Frick},
  \citenamefont {M\"uther}, \citenamefont {Rios}, \citenamefont {Polls},\ and\
  \citenamefont {Ramos}}]{Frick:2005}%
  \BibitemOpen
  \bibfield  {author} {\bibinfo {author} {\bibfnamefont {T.}~\bibnamefont
  {Frick}}, \bibinfo {author} {\bibfnamefont {H.}~\bibnamefont {M\"uther}},
  \bibinfo {author} {\bibfnamefont {A.}~\bibnamefont {Rios}}, \bibinfo {author}
  {\bibfnamefont {A.}~\bibnamefont {Polls}}, \ and\ \bibinfo {author}
  {\bibfnamefont {A.}~\bibnamefont {Ramos}},\ }\href {\doibase
  10.1103/PhysRevC.71.014313} {\bibfield  {journal} {\bibinfo  {journal} {Phys.
  Rev. C}\ }\textbf {\bibinfo {volume} {71}},\ \bibinfo {pages} {014313}
  (\bibinfo {year} {2005})}\BibitemShut {NoStop}%
\bibitem [{\citenamefont {Rios}\ \emph {et~al.}(2009)\citenamefont {Rios},
  \citenamefont {Polls},\ and\ \citenamefont {Dickhoff}}]{Rios:2009}%
  \BibitemOpen
  \bibfield  {author} {\bibinfo {author} {\bibfnamefont {A.}~\bibnamefont
  {Rios}}, \bibinfo {author} {\bibfnamefont {A.}~\bibnamefont {Polls}}, \ and\
  \bibinfo {author} {\bibfnamefont {W.~H.}\ \bibnamefont {Dickhoff}},\ }\href
  {\doibase 10.1103/PhysRevC.79.064308} {\bibfield  {journal} {\bibinfo
  {journal} {Phys. Rev. C}\ }\textbf {\bibinfo {volume} {79}},\ \bibinfo
  {pages} {064308} (\bibinfo {year} {2009})}\BibitemShut {NoStop}%
\bibitem [{\citenamefont {Rios}\ \emph {et~al.}(2014)\citenamefont {Rios},
  \citenamefont {Polls},\ and\ \citenamefont {Dickhoff}}]{Rios:2014}%
  \BibitemOpen
  \bibfield  {author} {\bibinfo {author} {\bibfnamefont {A.}~\bibnamefont
  {Rios}}, \bibinfo {author} {\bibfnamefont {A.}~\bibnamefont {Polls}}, \ and\
  \bibinfo {author} {\bibfnamefont {W.~H.}\ \bibnamefont {Dickhoff}},\ }\href
  {\doibase 10.1103/PhysRevC.89.044303} {\bibfield  {journal} {\bibinfo
  {journal} {Phys. Rev. C}\ }\textbf {\bibinfo {volume} {89}},\ \bibinfo
  {pages} {044303} (\bibinfo {year} {2014})}\BibitemShut {NoStop}%
\bibitem [{\citenamefont {Wiringa}\ \emph {et~al.}(2014)\citenamefont
  {Wiringa}, \citenamefont {Schiavilla}, \citenamefont {Pieper},\ and\
  \citenamefont {Carlson}}]{Wiringa:2014}%
  \BibitemOpen
  \bibfield  {author} {\bibinfo {author} {\bibfnamefont {R.~B.}\ \bibnamefont
  {Wiringa}}, \bibinfo {author} {\bibfnamefont {R.}~\bibnamefont {Schiavilla}},
  \bibinfo {author} {\bibfnamefont {S.~C.}\ \bibnamefont {Pieper}}, \ and\
  \bibinfo {author} {\bibfnamefont {J.}~\bibnamefont {Carlson}},\ }\href
  {\doibase 10.1103/PhysRevC.89.024305} {\bibfield  {journal} {\bibinfo
  {journal} {Phys. Rev. C}\ }\textbf {\bibinfo {volume} {89}},\ \bibinfo
  {pages} {024305} (\bibinfo {year} {2014})}\BibitemShut {NoStop}%
\bibitem [{\citenamefont {Audi}\ \emph {et~al.}(2003)\citenamefont {Audi},
  \citenamefont {Wapstra},\ and\ \citenamefont {Thibault}}]{AME:2003}%
  \BibitemOpen
  \bibfield  {author} {\bibinfo {author} {\bibfnamefont {G.}~\bibnamefont
  {Audi}}, \bibinfo {author} {\bibfnamefont {A.}~\bibnamefont {Wapstra}}, \
  and\ \bibinfo {author} {\bibfnamefont {C.}~\bibnamefont {Thibault}},\ }\href
  {\doibase https://doi.org/10.1016/j.nuclphysa.2003.11.003} {\bibfield
  {journal} {\bibinfo  {journal} {Nucl. Phys. A}\ }\textbf {\bibinfo {volume}
  {729}},\ \bibinfo {pages} {337 } (\bibinfo {year} {2003})},\ \bibinfo {note}
  {the 2003 NUBASE and Atomic Mass Evaluations}\BibitemShut {NoStop}%
\bibitem [{\citenamefont {Becker}\ \emph {et~al.}(2018)\citenamefont {Becker},
  \citenamefont {Bucoveanu}, \citenamefont {Grzesik}, \citenamefont {Imai},
  \citenamefont {Kempf}, \citenamefont {Molitor}, \citenamefont {Tyukin},
  \citenamefont {Zimmermann}, \citenamefont {Armstrong}, \citenamefont
  {Aulenbacher}, \citenamefont {Baunack}, \citenamefont {Beminiwattha},
  \citenamefont {Berger}, \citenamefont {Bernhard}, \citenamefont {Brogna},
  \citenamefont {Capozza}, \citenamefont {Covrig~Dusa}, \citenamefont
  {Deconinck}, \citenamefont {Diefenbach}, \citenamefont {Dunne}, \citenamefont
  {Erler}, \citenamefont {Gal}, \citenamefont {Gericke}, \citenamefont
  {Gl{\"a}ser}, \citenamefont {Gorchtein}, \citenamefont {Gou}, \citenamefont
  {Gradl}, \citenamefont {Imai}, \citenamefont {Kumar}, \citenamefont {Maas},
  \citenamefont {Mammei}, \citenamefont {Pan}, \citenamefont {Pandey},
  \citenamefont {Paschke}, \citenamefont {Peri{\'{c}}}, \citenamefont {Pitt},
  \citenamefont {Rahman}, \citenamefont {Riordan}, \citenamefont
  {Rodr{\'i}guez~Pi{\~{n}}eiro}, \citenamefont {Sfienti}, \citenamefont
  {Sorokin}, \citenamefont {Souder}, \citenamefont {Spiesberger}, \citenamefont
  {Thiel}, \citenamefont {Tyukin},\ and\ \citenamefont {Weitzel}}]{Becker2018}%
  \BibitemOpen
  \bibfield  {author} {\bibinfo {author} {\bibfnamefont {D.}~\bibnamefont
  {Becker}}, \bibinfo {author} {\bibfnamefont {R.}~\bibnamefont {Bucoveanu}},
  \bibinfo {author} {\bibfnamefont {C.}~\bibnamefont {Grzesik}}, \bibinfo
  {author} {\bibfnamefont {K.}~\bibnamefont {Imai}}, \bibinfo {author}
  {\bibfnamefont {R.}~\bibnamefont {Kempf}}, \bibinfo {author} {\bibfnamefont
  {M.}~\bibnamefont {Molitor}}, \bibinfo {author} {\bibfnamefont
  {A.}~\bibnamefont {Tyukin}}, \bibinfo {author} {\bibfnamefont
  {M.}~\bibnamefont {Zimmermann}}, \bibinfo {author} {\bibfnamefont
  {D.}~\bibnamefont {Armstrong}}, \bibinfo {author} {\bibfnamefont
  {K.}~\bibnamefont {Aulenbacher}}, \bibinfo {author} {\bibfnamefont
  {S.}~\bibnamefont {Baunack}}, \bibinfo {author} {\bibfnamefont
  {R.}~\bibnamefont {Beminiwattha}}, \bibinfo {author} {\bibfnamefont
  {N.}~\bibnamefont {Berger}}, \bibinfo {author} {\bibfnamefont
  {P.}~\bibnamefont {Bernhard}}, \bibinfo {author} {\bibfnamefont
  {A.}~\bibnamefont {Brogna}}, \bibinfo {author} {\bibfnamefont
  {L.}~\bibnamefont {Capozza}}, \bibinfo {author} {\bibfnamefont
  {S.}~\bibnamefont {Covrig~Dusa}}, \bibinfo {author} {\bibfnamefont
  {W.}~\bibnamefont {Deconinck}}, \bibinfo {author} {\bibfnamefont
  {J.}~\bibnamefont {Diefenbach}}, \bibinfo {author} {\bibfnamefont
  {J.}~\bibnamefont {Dunne}}, \bibinfo {author} {\bibfnamefont
  {J.}~\bibnamefont {Erler}}, \bibinfo {author} {\bibfnamefont
  {C.}~\bibnamefont {Gal}}, \bibinfo {author} {\bibfnamefont {M.}~\bibnamefont
  {Gericke}}, \bibinfo {author} {\bibfnamefont {B.}~\bibnamefont {Gl{\"a}ser}},
  \bibinfo {author} {\bibfnamefont {M.}~\bibnamefont {Gorchtein}}, \bibinfo
  {author} {\bibfnamefont {B.}~\bibnamefont {Gou}}, \bibinfo {author}
  {\bibfnamefont {W.}~\bibnamefont {Gradl}}, \bibinfo {author} {\bibfnamefont
  {Y.}~\bibnamefont {Imai}}, \bibinfo {author} {\bibfnamefont {K.~S.}\
  \bibnamefont {Kumar}}, \bibinfo {author} {\bibfnamefont {F.}~\bibnamefont
  {Maas}}, \bibinfo {author} {\bibfnamefont {J.}~\bibnamefont {Mammei}},
  \bibinfo {author} {\bibfnamefont {J.}~\bibnamefont {Pan}}, \bibinfo {author}
  {\bibfnamefont {P.}~\bibnamefont {Pandey}}, \bibinfo {author} {\bibfnamefont
  {K.}~\bibnamefont {Paschke}}, \bibinfo {author} {\bibfnamefont
  {I.}~\bibnamefont {Peri{\'{c}}}}, \bibinfo {author} {\bibfnamefont
  {M.}~\bibnamefont {Pitt}}, \bibinfo {author} {\bibfnamefont {S.}~\bibnamefont
  {Rahman}}, \bibinfo {author} {\bibfnamefont {S.}~\bibnamefont {Riordan}},
  \bibinfo {author} {\bibfnamefont {D.}~\bibnamefont
  {Rodr{\'i}guez~Pi{\~{n}}eiro}}, \bibinfo {author} {\bibfnamefont
  {C.}~\bibnamefont {Sfienti}}, \bibinfo {author} {\bibfnamefont
  {I.}~\bibnamefont {Sorokin}}, \bibinfo {author} {\bibfnamefont
  {P.}~\bibnamefont {Souder}}, \bibinfo {author} {\bibfnamefont
  {H.}~\bibnamefont {Spiesberger}}, \bibinfo {author} {\bibfnamefont
  {M.}~\bibnamefont {Thiel}}, \bibinfo {author} {\bibfnamefont
  {V.}~\bibnamefont {Tyukin}}, \ and\ \bibinfo {author} {\bibfnamefont
  {Q.}~\bibnamefont {Weitzel}},\ }\href {\doibase 10.1140/epja/i2018-12611-6}
  {\bibfield  {journal} {\bibinfo  {journal} {Eur. Phys. J. A}\ }\textbf
  {\bibinfo {volume} {54}},\ \bibinfo {pages} {208} (\bibinfo {year}
  {2018})}\BibitemShut {NoStop}%
\bibitem [{\citenamefont {Horowitz}\ \emph {et~al.}(2014)\citenamefont
  {Horowitz}, \citenamefont {Kumar},\ and\ \citenamefont
  {Michaels}}]{Horowitz14}%
  \BibitemOpen
  \bibfield  {author} {\bibinfo {author} {\bibfnamefont {C.~J.}\ \bibnamefont
  {Horowitz}}, \bibinfo {author} {\bibfnamefont {K.~S.}\ \bibnamefont {Kumar}},
  \ and\ \bibinfo {author} {\bibfnamefont {R.}~\bibnamefont {Michaels}},\
  }\href@noop {} {\bibfield  {journal} {\bibinfo  {journal} {Eur. Phys. J. A}\
  }\textbf {\bibinfo {volume} {50}},\ \bibinfo {pages} {48} (\bibinfo {year}
  {2014})}\BibitemShut {NoStop}%
\bibitem [{\citenamefont {Hagen}\ \emph {et~al.}(2016)\citenamefont {Hagen},
  \citenamefont {Ekstr\"{o}m}, \citenamefont {Forss\'{e}n}, \citenamefont
  {Jansen}, \citenamefont {Nazarewicz}, \citenamefont {Papenbrock},
  \citenamefont {Wendt}, \citenamefont {Bacca}, \citenamefont {Barnea},
  \citenamefont {Carlsson}, \citenamefont {Drischler}, \citenamefont {Hebeler},
  \citenamefont {Hjorth-Jenson}, \citenamefont {Miorelli}, \citenamefont
  {Orlandini}, \citenamefont {Schwenk},\ and\ \citenamefont
  {Simonis}}]{Hagen:2016}%
  \BibitemOpen
  \bibfield  {author} {\bibinfo {author} {\bibfnamefont {G.}~\bibnamefont
  {Hagen}}, \bibinfo {author} {\bibfnamefont {A.}~\bibnamefont {Ekstr\"{o}m}},
  \bibinfo {author} {\bibfnamefont {C.}~\bibnamefont {Forss\'{e}n}}, \bibinfo
  {author} {\bibfnamefont {G.~R.}\ \bibnamefont {Jansen}}, \bibinfo {author}
  {\bibfnamefont {W.}~\bibnamefont {Nazarewicz}}, \bibinfo {author}
  {\bibfnamefont {T.}~\bibnamefont {Papenbrock}}, \bibinfo {author}
  {\bibfnamefont {K.~A.}\ \bibnamefont {Wendt}}, \bibinfo {author}
  {\bibfnamefont {S.}~\bibnamefont {Bacca}}, \bibinfo {author} {\bibfnamefont
  {N.}~\bibnamefont {Barnea}}, \bibinfo {author} {\bibfnamefont
  {B.}~\bibnamefont {Carlsson}}, \bibinfo {author} {\bibfnamefont
  {C.}~\bibnamefont {Drischler}}, \bibinfo {author} {\bibfnamefont
  {K.}~\bibnamefont {Hebeler}}, \bibinfo {author} {\bibfnamefont
  {M.}~\bibnamefont {Hjorth-Jenson}}, \bibinfo {author} {\bibfnamefont
  {M.}~\bibnamefont {Miorelli}}, \bibinfo {author} {\bibfnamefont
  {G.}~\bibnamefont {Orlandini}}, \bibinfo {author} {\bibfnamefont
  {A.}~\bibnamefont {Schwenk}}, \ and\ \bibinfo {author} {\bibfnamefont
  {J.}~\bibnamefont {Simonis}},\ }\href@noop {} {\bibfield  {journal} {\bibinfo
   {journal} {Nature Phys.}\ }\textbf {\bibinfo {volume} {12}},\ \bibinfo
  {pages} {186} (\bibinfo {year} {2016})}\BibitemShut {NoStop}%
\bibitem [{\citenamefont {Mammei}\ \emph {et~al.}(2013)\citenamefont {Mammei}
  \emph {et~al.}}]{CREX13}%
  \BibitemOpen
  \bibfield  {author} {\bibinfo {author} {\bibfnamefont {J.}~\bibnamefont
  {Mammei}} \emph {et~al.},\ }\href@noop {} {\enquote {\bibinfo {title}
  {{CREX}: Parity-violating measurement of the weak charge distribution of
  ${}^{48}${Ca} to 0.02 fm accuracy},}\ }\bibinfo {howpublished}
  {http://hallaweb.jlab.org/parity/prex/} (\bibinfo {year} {2013})\BibitemShut
  {NoStop}%
\bibitem [{\citenamefont {Souder}\ \emph {et~al.}(2011)\citenamefont {Souder}
  \emph {et~al.}}]{PREX-II}%
  \BibitemOpen
  \bibfield  {author} {\bibinfo {author} {\bibfnamefont {P.~A.}\ \bibnamefont
  {Souder}} \emph {et~al.},\ }\href@noop {} {\enquote {\bibinfo {title}
  {{PREX-II}: Precision parity-violating measurement of the neutron skin of
  lead},}\ }\bibinfo {howpublished} {http://hallaweb.jlab.org/parity/prex/}
  (\bibinfo {year} {2011})\BibitemShut {NoStop}%
\bibitem [{\citenamefont {Thiel}\ \emph {et~al.}(2019)\citenamefont {Thiel},
  \citenamefont {Sfienti}, \citenamefont {Piekarewicz}, \citenamefont
  {Horowitz},\ and\ \citenamefont {Vanderhaeghen}}]{Thiel:2019}%
  \BibitemOpen
  \bibfield  {author} {\bibinfo {author} {\bibfnamefont {M.}~\bibnamefont
  {Thiel}}, \bibinfo {author} {\bibfnamefont {C.}~\bibnamefont {Sfienti}},
  \bibinfo {author} {\bibfnamefont {J.}~\bibnamefont {Piekarewicz}}, \bibinfo
  {author} {\bibfnamefont {C.~J.}\ \bibnamefont {Horowitz}}, \ and\ \bibinfo
  {author} {\bibfnamefont {M.}~\bibnamefont {Vanderhaeghen}},\ }\href@noop {}
  {\bibfield  {journal} {\bibinfo  {journal} {J. Phys. G: Nucl. Part. Phys.}\
  }\textbf {\bibinfo {volume} {46}},\ \bibinfo {pages} {093003} (\bibinfo
  {year} {2019})}\BibitemShut {NoStop}%
\bibitem [{\citenamefont {Abbott}\ \emph {et~al.}(2017)\citenamefont {Abbott}
  \emph {et~al.}}]{LIGO:2017}%
  \BibitemOpen
  \bibfield  {author} {\bibinfo {author} {\bibfnamefont {B.~P.}\ \bibnamefont
  {Abbott}} \emph {et~al.} (\bibinfo {collaboration} {LIGO Scientific
  Collaboration and Virgo Collaboration}),\ }\href {\doibase
  10.1103/PhysRevLett.119.161101} {\bibfield  {journal} {\bibinfo  {journal}
  {Phys. Rev. Lett.}\ }\textbf {\bibinfo {volume} {119}},\ \bibinfo {pages}
  {161101} (\bibinfo {year} {2017})}\BibitemShut {NoStop}%
\bibitem [{\citenamefont {Baye}(2015)}]{Baye_review}%
  \BibitemOpen
  \bibfield  {author} {\bibinfo {author} {\bibfnamefont {D.}~\bibnamefont
  {Baye}},\ }\href {\doibase https://doi.org/10.1016/j.physrep.2014.11.006}
  {\bibfield  {journal} {\bibinfo  {journal} {Phys. Rep.}\ }\textbf {\bibinfo
  {volume} {565}},\ \bibinfo {pages} {1 } (\bibinfo {year} {2015})}\BibitemShut
  {NoStop}%
\end{thebibliography}%

\end{document}